\documentclass[]{pasj01} 
\Received{}
\Accepted{}
\usepackage[switch,mathlines]{lineno}
\usepackage{graphicx}	
\usepackage{bm}  
\usepackage{color} 
\usepackage[T1]{fontenc} 
\usepackage[switch,mathlines]{lineno} 
\DeclareRobustCommand{\VAN}[3]{#2}
\let\VANthebibliography\thebibliography
\def\thebibliography{\DeclareRobustCommand{\VAN}[3]{##3}\VANthebibliography}  

\def\be{\begin{equation}}\def\ee{\end{equation}}
\def\vlsr{v_{\rm LSR}} \def\Msun{M_\odot} \def\deg{^\circ}  
\def\co{$^{12}$CO } \def\coth{$^{13}$CO } \def\Xco{X_{\rm CO}}   \def\Tb{T_{\rm B}} 
\def\Htwo{H$_2$ }   \def\Kkms{K km s$^{-1}}   \def\kms{km s$^{-1}$} \def\Ico{I_{\rm CO}}  \def\Kkms{K \kms }\def\mH{m_{\rm H}} 
 \def\Ico{I_{\rm CO}} \def\htwo{H$_2$} \def\Tb{T_{\rm B}}   \def\mH{m_{\rm H}}     \def\ekms{{\rm km\ s^{-1}}}  \def\epc{{\rm pc} } 
 \def\apj{ApJ} \def\aap{AA} \def\mnras{MNRAS} \def\pasj{PASJ} \def\aj{AJ}
 \def\xcounit{H$_2$ cm $^{-2}$ [K km s$^{-1}]^{-1}$}

\def\Msun{M_{\odot \hskip-5.2pt \bullet}}    \def\kms{km s$^{-1}$}  \def\deg{^\circ}   \def\Htwo{H$_2$\ }

\def\bc{\begin{center}}\def\ec{\end{center}}

\def\red{}

\def\ekms{{\rm km\ s^{-1}}}
\def\vexpa{v_{\rm expa}} 
\def\rbub{r_{\rm b}}
\def\x{\times}\def\xfour{\times 10^4}\def\xfifty{\times 10^{50}}\def\xmtwe{\times 10^{-20}}
 
\def\Xgc{X_{\rm CO;GC}} 
\def\Xbri{X_{\rm CO;Bri}}
\def\Mbrixco{M_{\rm Bri;Xco}}
\def\Mbrivir{M_{\rm Bri;vir}}
\def\Mbub{M_{\rm bub}}
\def\rbub{r_{\rm bub}}
\def\rbri{r_{\rm Bri}}
\def\revtwo{}

\begin{document}   

\title{Dark supernova remnant buried in the Galactic-Centre "Brick" G0.253+0.016 revealed by an expanding CO-line bubble}
\author{Yoshiaki \textsc{Sofue}\altaffilmark{1} }
\altaffiltext{1}{Institute of Astronomy, The University of Tokyo, Mitaka, Tokyo 183-0015, Japan}
\email{sofue@ioa.s.u-tokyo.ac.jp}

 
\maketitle 

\begin{abstract}     
We performed a \co- and \coth-line study of the "Brick" (G0.253+0.016) in the Galactic Centre (GC) by analyzing archival data obtained with the Nobeyama 45-m telescope.
We present kinematics and molecular gas distributions in the longitude-velocity diagram, and suggest that the Brick is located along the GC Arm I in the central molecular zone (CMZ), which yields a distance from the Sun of 8 kpc and galacto-centric distance of 0.2 kpc.
The major and minor-axis diameters of the Brick are $D_x\times D_y=8.4 \epc \times 4.1 \epc$ at position angle of $40\deg$ and $130\deg$, respectively, and the scale radius is $\rbri=\sqrt{D_x D_y}=2.96 \epc$.
The molecular mass inferred from the \co-line integrated intensity is $\Mbrixco\sim 5.1\times 10^4 \Msun$ for a conversion factor $\Xgc=1.0\times 10^{20}$ \xcounit.
On the other hand, the dynamical (Virial) mass for the measured velocity dispersion of $\sigma_v=10.0 \ekms$ is calculated to be $\Mbrivir\sim 6.8 \times 10^4\Msun$, which yields a new conversion factor of $\Xbri =1.3\times 10^{20}$ \xcounit. 
The Brick's center has a cavity surrounded by a spherical molecular bubble of radius $\rbub=1.85$ pc and mass $\sim 1.7\times 10^4\Msun$ expanding at $v_{\rm exp}\simeq 10 \ \ekms$ with kinetic energy of $E_0\sim 1.7\times 10^{49}$ erg.
If the bubble is approximated by an adiabatic spherical shock wave, the age is estimated to be $t\sim 2/5 r_{\rm bub}/v_{\rm exp}\sim 7.2\times 10^4$ y. 
Neither non-thermal radio structures nor thermal radio emission indicative of HII region are found in the archival data from the MeerKAT.
We suggest that the molecular bubble will be a dark supernova remnant buried in the Brick, which, therefore, has experienced in the past ($\sim 0.1$ Myr ago) massive-star formation with a supernova explosion.
\end{abstract}    



\section{Introduction}  

The "Brick" (G0.253+0.016, M0.25+0.01) near the Galactic Centre (GC) is a dense dust cloud detected in the submillimeter emission \citep{gue83,lis94a,joh14}.
Submillimeter photometry showed a molecular gas column density of $\sim 10^{23}$ \htwo\  cm$^{-2}$ and a mass of $\sim 1.5\times 10^5\Msun$ \citep{lon12}.
It exhibits severe extinction at infrared wavelengths, silhouetted against the central stellar disk and bulge \citep{hen19,gin23}.
The cloud is detected by a molecular line with radial velocity $\vlsr \sim 30 \ekms$ \citep{joh14,lis94b}.
Interferometric observations with ALMA (Atacama Large Millimeter/Submillimeter Array) revealed a bubble/cavity structure near the center of the brick, consisting of a semi-loop with numerous arcs concentric with the center of the bubble \citep{hig14,hen22}.
Filament and turbulent structures associated with magnetic fields are also highlighted \citep{fed16}.

There have been two models to explain the bubble/arc structure:
One idea is that it is a cavity produced by a collision of a compact molecular cloud at high speed from high-latitude direction \citep{hig14}.
The other attributes the bubble to stellar feedback such as a wind from young stars \citep{joh14,hen19,hen22}. 
While signature of star formation has been found by detection of maser lines and outflow from young stars \citep{lis94a,lon12,wal21}, no clear evidence is yet reported of HII regions indicative of massive-star formation \citep{and14,wen23}.

Infrared photometry of the bulge stars and extinction study indicated a distance of $\sim 7$ kpc \citep{lon12,zoc21}, locating the cloud about 1 kpc in front of the GC.
On the other hand, color-magnitude study of bulge stars \citep{nog21} and extinction study of GC stars with known proper motions \citep{mar22} suggested that the Brick is inside the central molecular zone (CMZ).

In this paper we study the kinematics and energetics of the Brick by analyzing the CO-line data from the GC survey with the Nobeyama 45-m mm-wave telescope \citep{tok19}.
We argue that the Brick is more likely to be associated with the CMZ located at a distance of $\sim 8$ kpc. 
\red{We derive the fundamental physical parameters such as the size, mass, kinetic and gravitational energies of the Brick based on the Nobeyama \co- and \coth-line data, which, because of the single-dish aperture, do not suffer from the missing-mass problem in interferometric measurments.
So, the present study will be complimentary to the current interferometer works in the sense that the present analysis provides with information more about physics that has not been explored in the current studies, while it does not add much information about the detailed morphology in the cloud.}

\red{We then focus on the expanding molecular bubble centered on $(l,b)=(0\deg.253,+0\deg.016)$ (G0.253+0.016), and present a new model attributing it to a dark supernova remnant (SNR) buried in the dense molecular Brick \citep{shu80,whe80,luc20,sof20,sof21} in order to explain the kinetic energy an order of magnitude greater than that estimated from the interferometer observations. }

\section{The Brick in the CO line}

\subsection{CO-line data}

We use the CO-line survey of the Galactic Center using the Nobeyama 45-m telescope \citep{tok19}, which cover the GC region for $1\deg.4\times 0\deg.8$ in the \co and \coth $(J=1-0)$-line emissions.
The full width of half maximum of the telescope beam was $15''$ and $16''$, respectively, corresponding to 0.60 and 0.64 pc at the GC distance of 8.2 kpc.
The data cubes have grid sizes of $7''.5\times 7''.5 \times 2 \ekms$.
Both bands were observed simultaneously, and the rms noise in the data cube was $\sim 0.2$ K.

\subsection{Intensity distribution: moment 0 map}

\begin{figure*}   
\begin{center}      
\includegraphics[width=12cm]{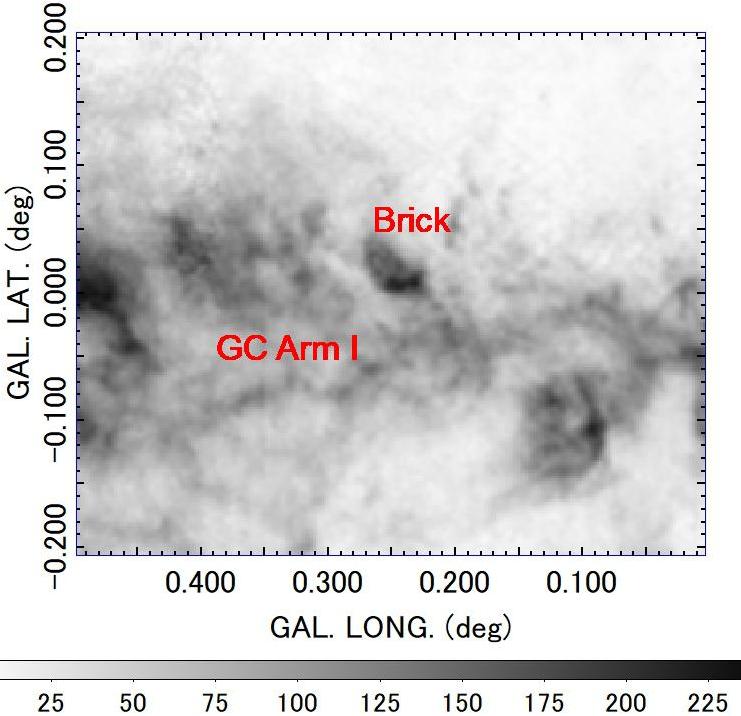}\\ 
\vskip 5mm 
\includegraphics[width=8cm]{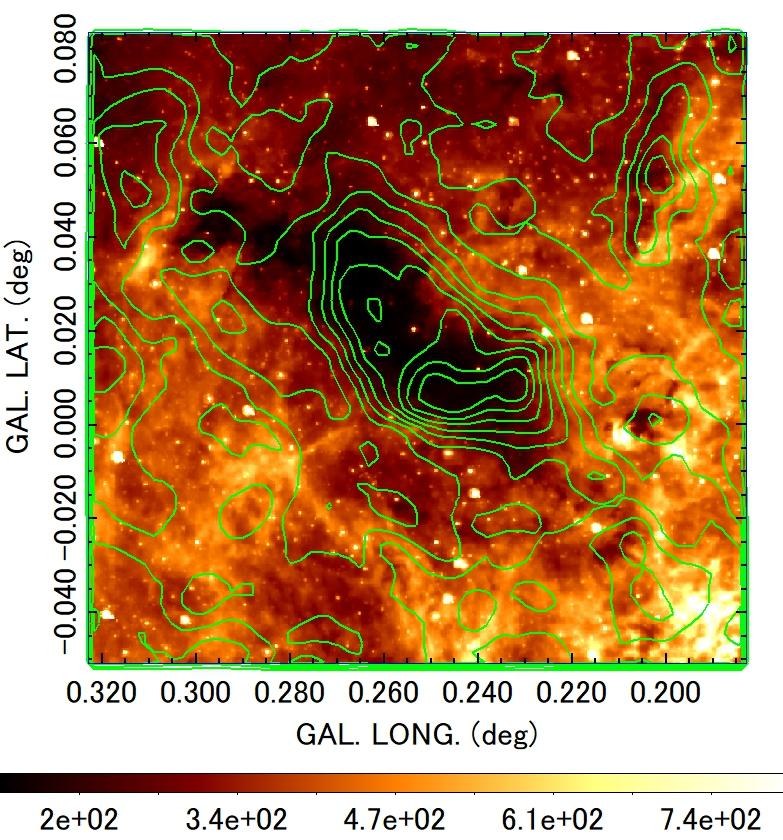} \hskip -5mm  
\includegraphics[width=9cm]{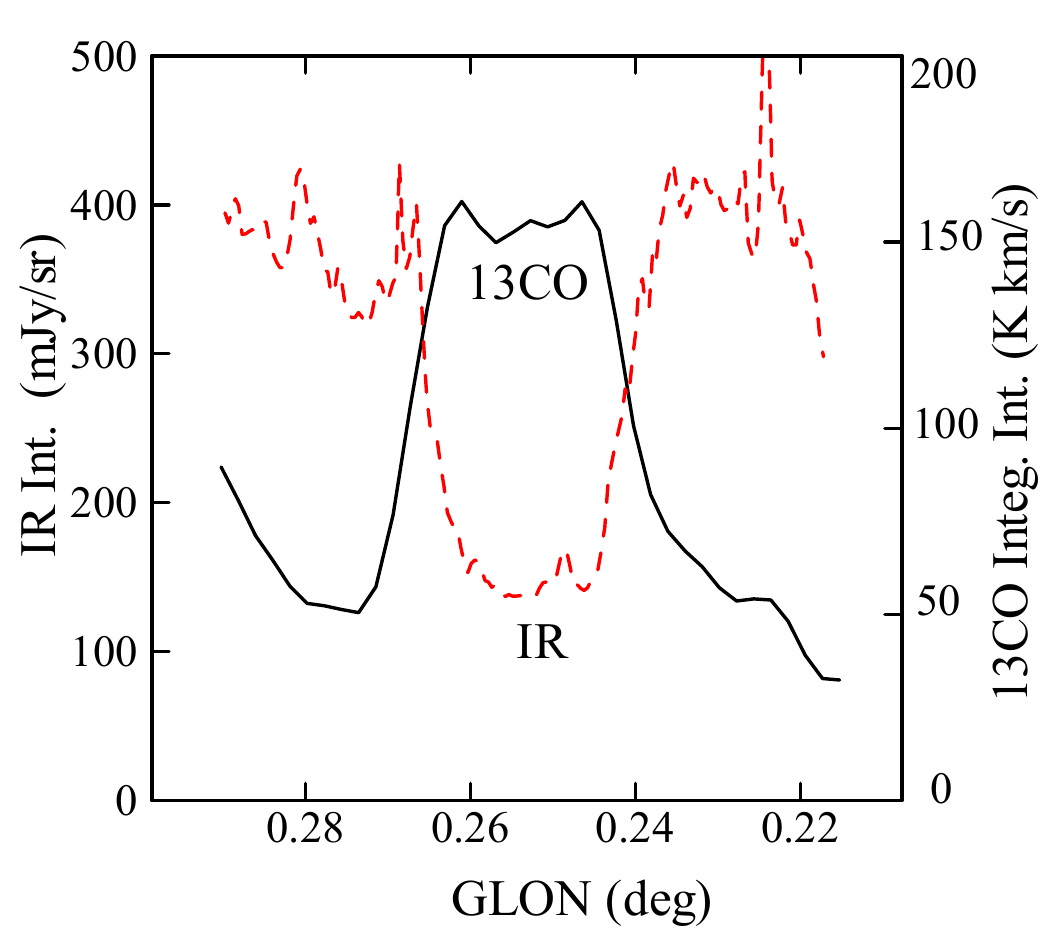}\\
(mJy sr$^{-1}$)\hskip 2cm ~~\\
\end{center}
\caption{[Top left] \coth\ intensity (moment 0) map integrated from $\vlsr=0$ to 60 \kms around the Brick.
[Top right] The Brick in \coth intensity by contours every 20 \Kkms overlaid on the Spitzer (GLIMPSE) 8 $\mu$m intensity map (grey scale in mJy/sr) \citep{chu09}, showing coincidence of CO and dust clouds.  
[Bottom] \coth intensity profiles (black line) and 8 $\mu$m (dashed line) perpendicular to the Brick's major axis across \red{$(l,b)=0\deg.253,+0\deg.016)$}.
}
\label{fig-m0+glimpse}	
\end{figure*}

Figure \ref{fig-m0+glimpse} shows integrated intensity (moment 0) maps in the \coth line emission between $\vlsr=0$ and 60 \kms for a $0\deg.5\times 0\deg.4$ (top panel) and $0\deg.16\times 0\deg.16$ (middle) regions around the Brick.
The middle panel is drawn by contours overlaid on the Spitzer (GLIMPSE) 8 $\mu$m intensity map in grey-scaling \citep{chu09}.
The map shows tight correlation of the CO cloud in emission with the dust Brick in silhouette against the stellar background of the central bulge and GC stellar disk. 
The bottom panel shows perpendicular cross sections of the Brick along the dashed line in the middle panel, showing plateaued intensity profiles both in infrared and \coth indicative of density cavity inside the brick as will be discussed later based on channel maps and position-velocity diagrams.

\begin{figure*}   
\begin{center}  
{\bf Moment 0 \co (green) + \coth (red)}\\
\includegraphics[width=8cm]{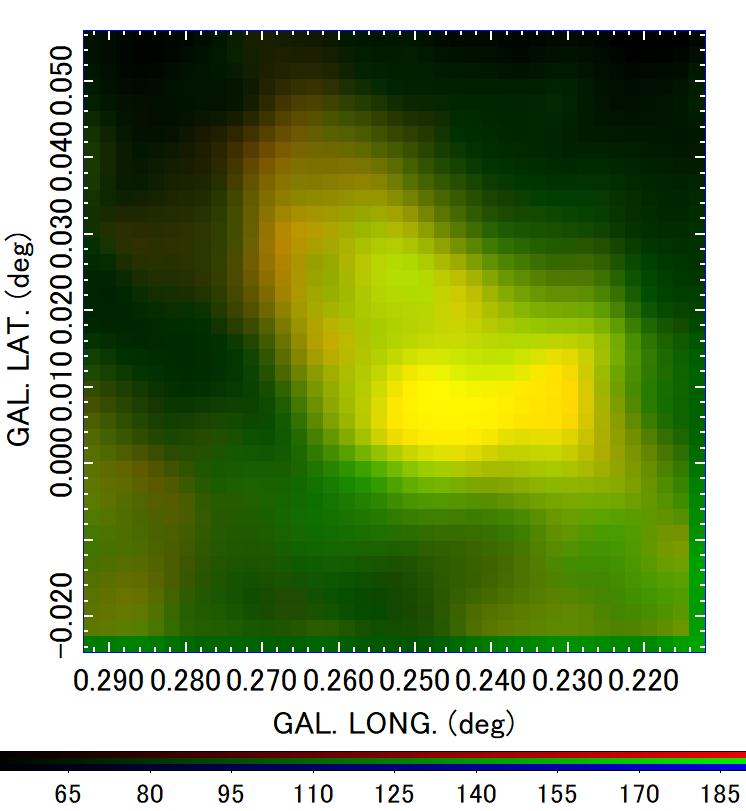}\\
(K \kms (\coth)) \\
\vskip 2mm
{\bf Moment 1 \hskip 6cm Moment 2}\\ 
\includegraphics[width=8cm]{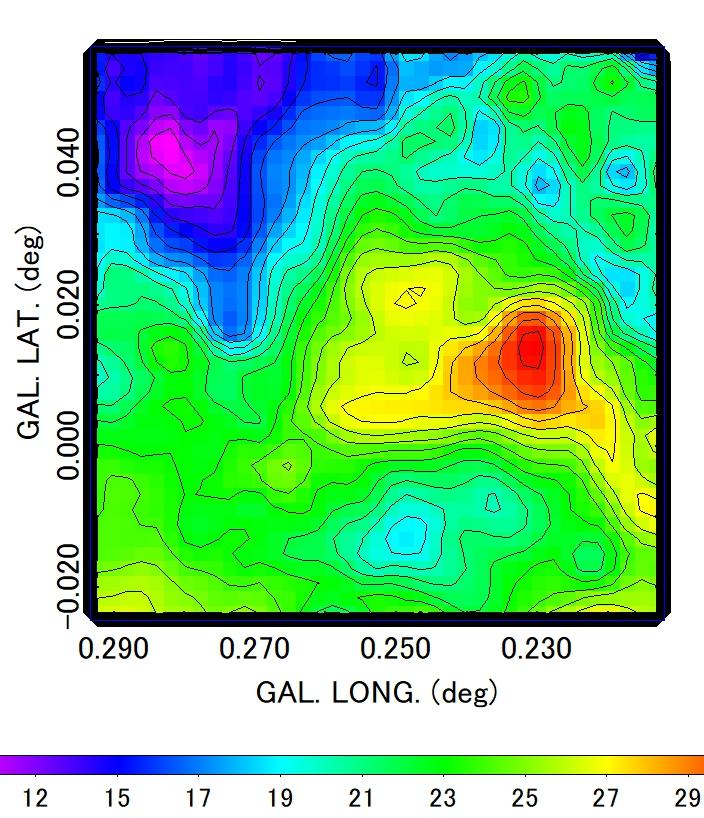} 
\includegraphics[width=8cm]{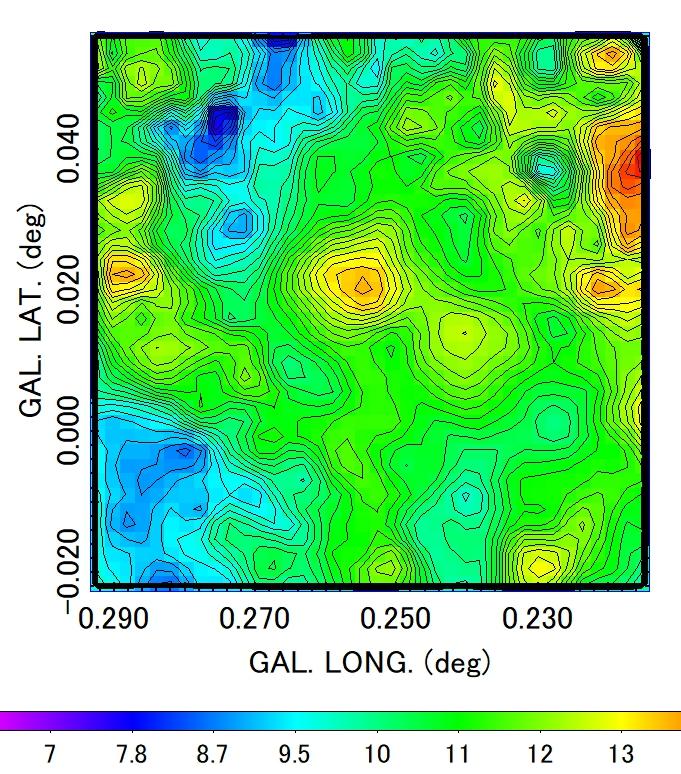}\\
\vskip -1mm (\kms) \hskip 8cm (\kms)\\ 
\end{center}
\caption{[Top] Composite map of integrated intensities from 0 to 60 \kms in the \co\ (green: from 250 to 1250 \Kkms) and \coth\ (red: 50 to 250) lines.
[bottom] Moment 1 (velocity field) and 2 (dispersion) in \kms.
}
\label{fig-mom}	
\end{figure*}

Figure \ref{fig-mom} shows integrated intensity (moment 0) maps of the \co and \coth line emissions of the Brick along with a velocity field (moment 1) and velocity dispersion map (moment 2).
The \co map shows a slightly smoother distribution than the map in \coth despite of the sharper beam because of the broader distribution as well as more sensitive detection of extended and diffuse gas clouds.
The bottom panel shows the cross section of the \coth intensity perpendicular to the Brick's major axis (position angle $\sim 130\deg$) across the map center of the middle panel.
The plateaued intensity distributions both in the infrared absorption and CO emission indicate a cavity inside the cloud, unless the brick is cubicle shaped.

\subsection{Velocity gradient and dispersion: moment 1 and 2 maps}

Figure \ref{fig-mom} shows moment 1 and 2 maps, showing the mean-velocity field and velocity dispersion distributions.
The moment-1 map indicates a gradual increase in the radial velocity from NE to SW, and will be discussed in section \ref{sec-rotation} in relation to the rotation of the cloud.
\revtwo{The moment 2 map shows a nearly uniform velocity dispersion at $\sim 10 \ \ekms$ over the Brick, except for the central region within the molecular cavity, where the dispersion attains a local maximum of $\sigma_v\sim 13$ \kms.
This central peak can be attributed to the maximum difference between the approaching and receding motions of the expanding shell across the center.}

\subsection{Channel maps}

\begin{figure*}   
\begin{center}  \includegraphics[width=14cm]{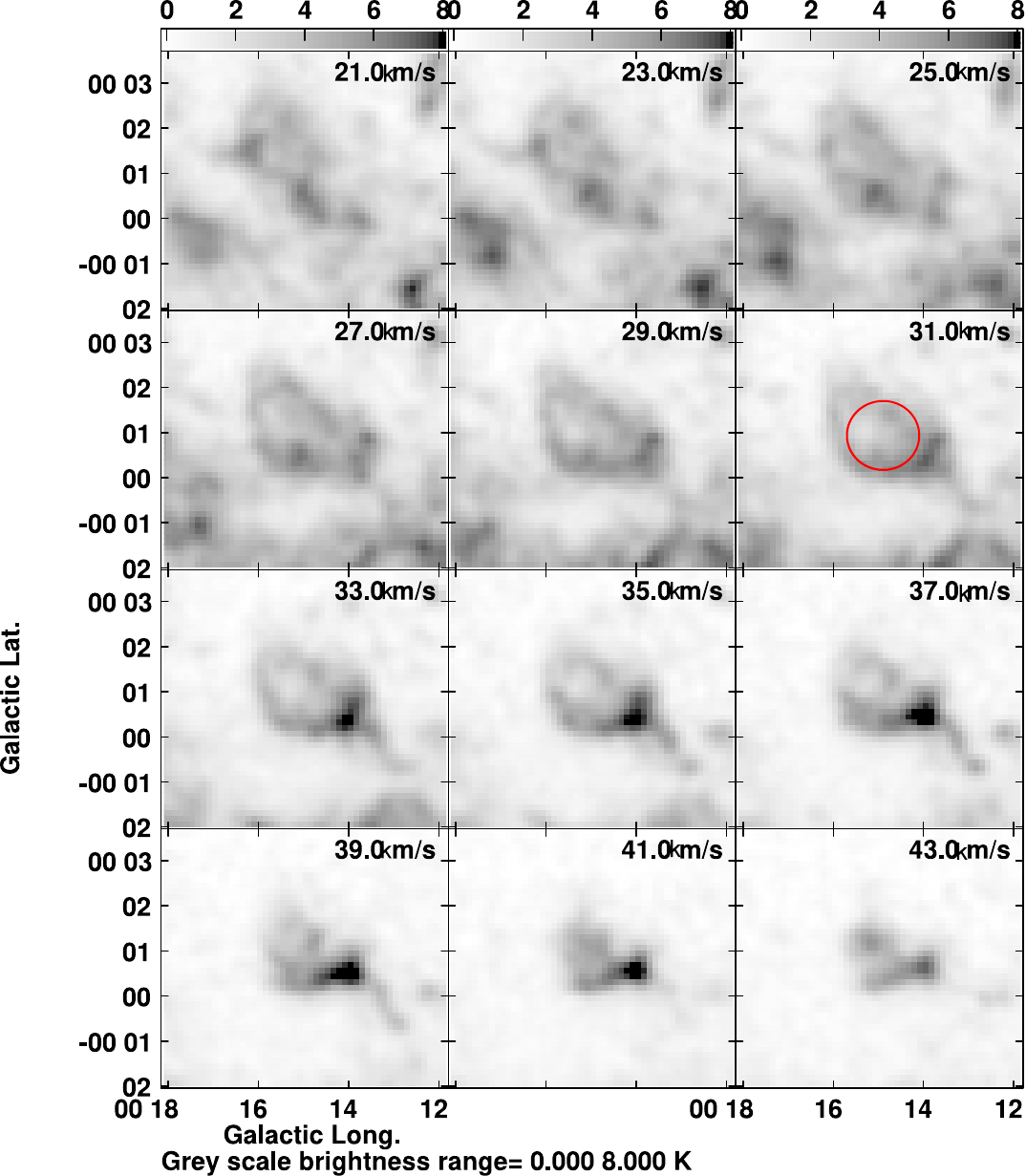} 
\end{center}
\caption{ Channel maps of \coth\ $\Tb$ (K by bar) of the Brick every 2 \kms interval of $\vlsr$. Note the ring-like distribution of the emission, indicating a shell structure.
\red{Red circle marks the bubble of radius 1.9 pc (at 8 kpc) centered on $(l,b)=(0\deg.253,+0\deg.016)$ with radius 1.9 pc.
Note tht coordinate values are in unit of $dd\deg\ mm'$.} }
\label{fig-chan}	
\end{figure*}

Figure \ref{fig-chan} shows channel maps of the brightness temperature $\Tb$ of the \coth emission around the Brick from $\vlsr=21$ to 43 \kms.
The maps show ring-like distributions of the brightness making a cavity in the central region.
The ring feature at velocity $\sim 30$ \kms slice is well fitted by a circle representing the cross section of a spherical bubble (shell) centered on \red{$(l,b)=(0\deg.253,+0\deg.016)$} with radius $\sim 1'.6$ (1.7 to 1.9 pc at $d=7.2$ to 8 kpc).
The bubble is associated with a peaky clump in the SW edge.
Combining with elliptical features in the position-velocity diagrams, this ring will be attributed to an expanding molecular shell (bubble) in the following subsections.

\begin{figure*} 
\begin{center} \includegraphics[width=12cm]{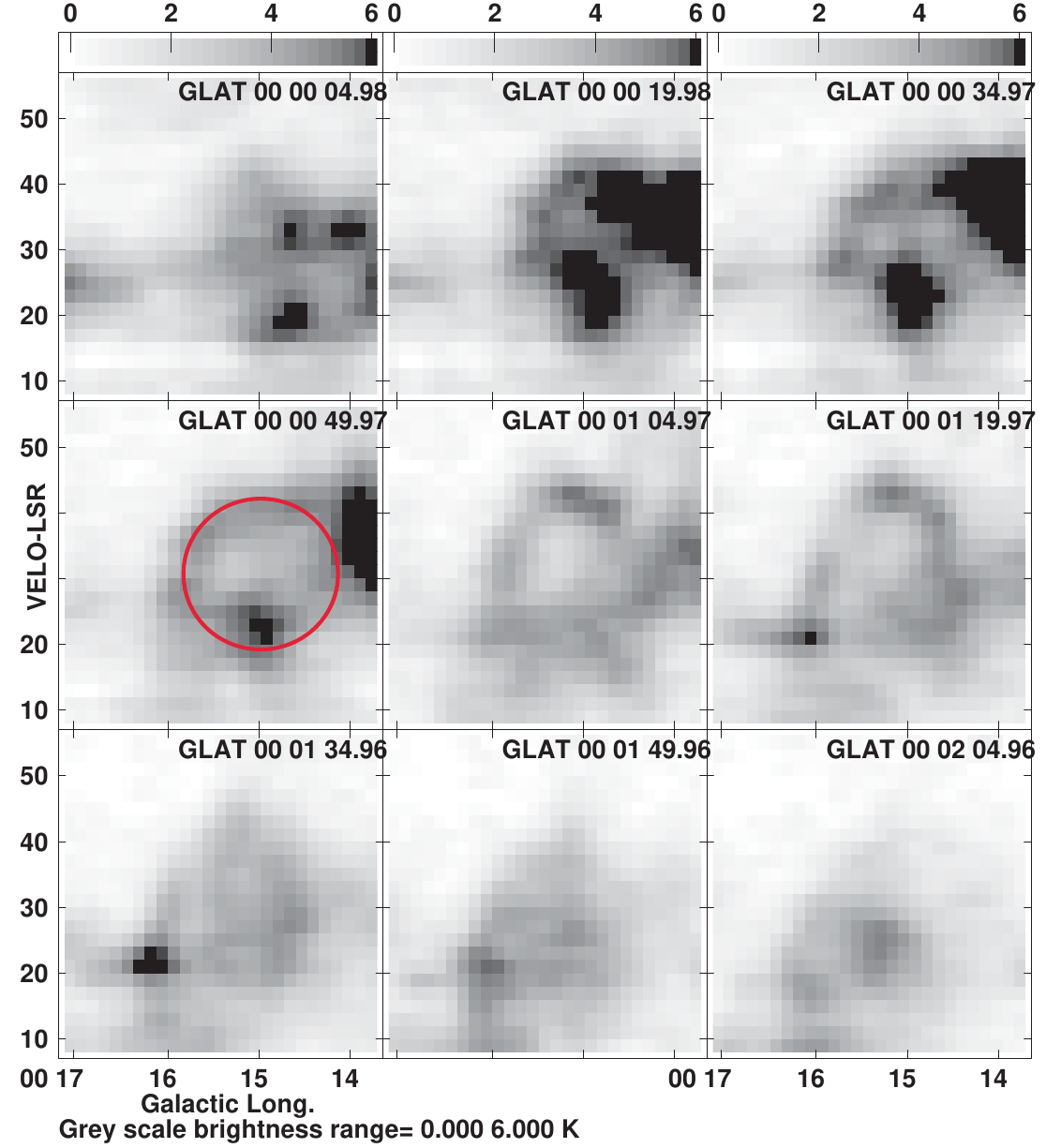}  
\end{center}
\caption{ LVDs at various latitudes.
Red circle traces the LV ellipse representing the expanding bubble at 10 \kms. 
\red{Note: Coordinate values are in unit of $dd\deg \ mm'$ and \kms and labels in unit of $dd\deg \ mm' \ ss''$.}} 
\label{fig-chan-lvd}	
\end{figure*} 
 
\begin{figure*}   
\begin{center}  
 \includegraphics[width=12cm]{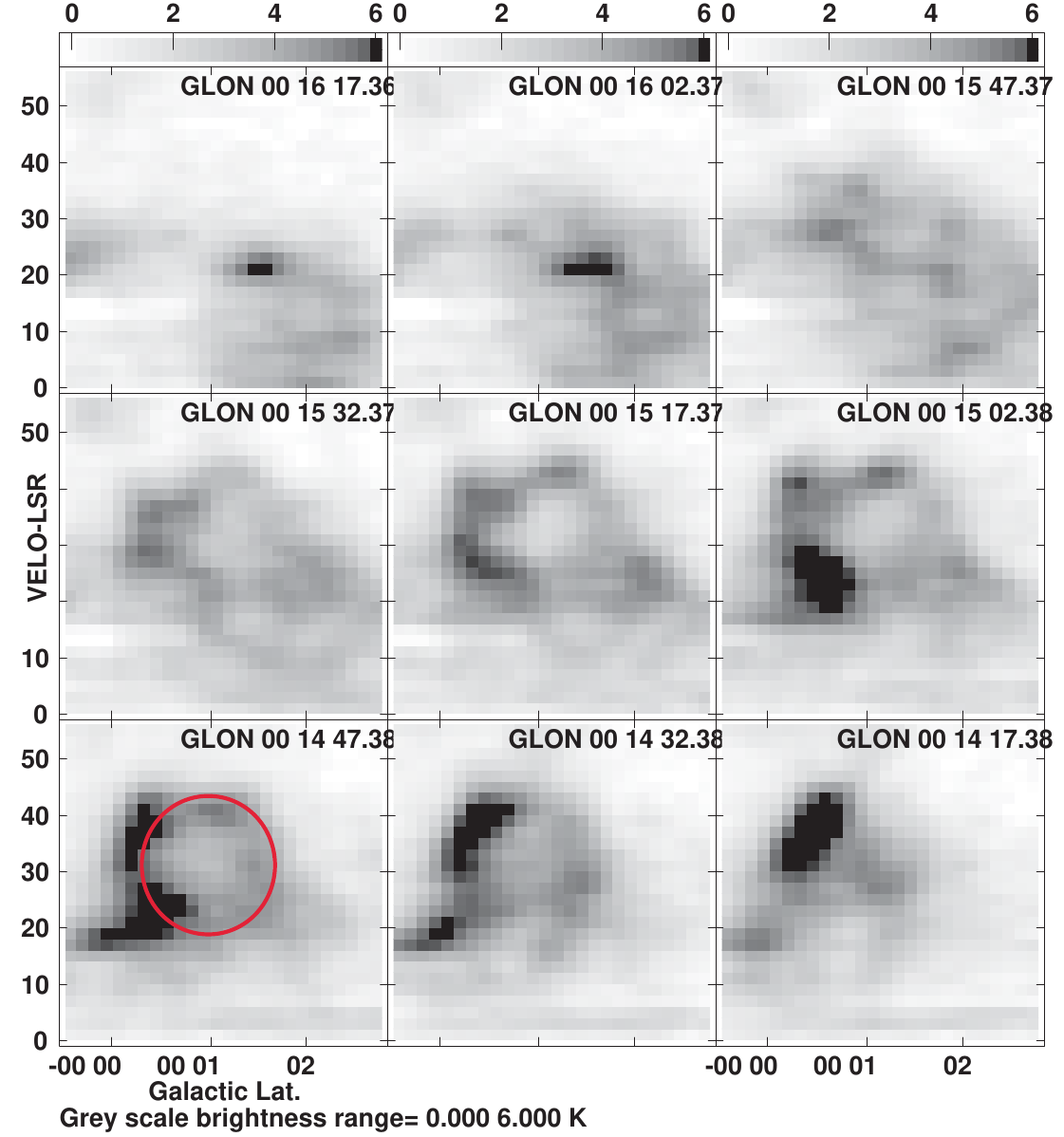} %
\end{center}
\caption{  BVDs. The red circle indicates the expanding bubble. 
\red{Note: Coordinate values are in unit of $dd\deg \ mm'$ and \kms and labels in unit of $dd\deg \ mm' \ ss''$.}
}
\label{fig-chan-bvd}	
\end{figure*} 
 
\subsection{Position-velocity diagrams}

In figures \ref{fig-chan-lvd} and \ref{fig-chan-bvd} we show  
longitude-velocity diagrams (LVD) sliced at various latitudes
and latitude-velocity digrams (BVD) sliced at various longitudes, respectively.
In both diagrams, elliptical features are recognized as marked by the red circles, indicating a bubble structure expanding at velocity of $\vexpa\sim 10$ \kms.

\begin{figure}   
\begin{center}     
\includegraphics[width=8cm]{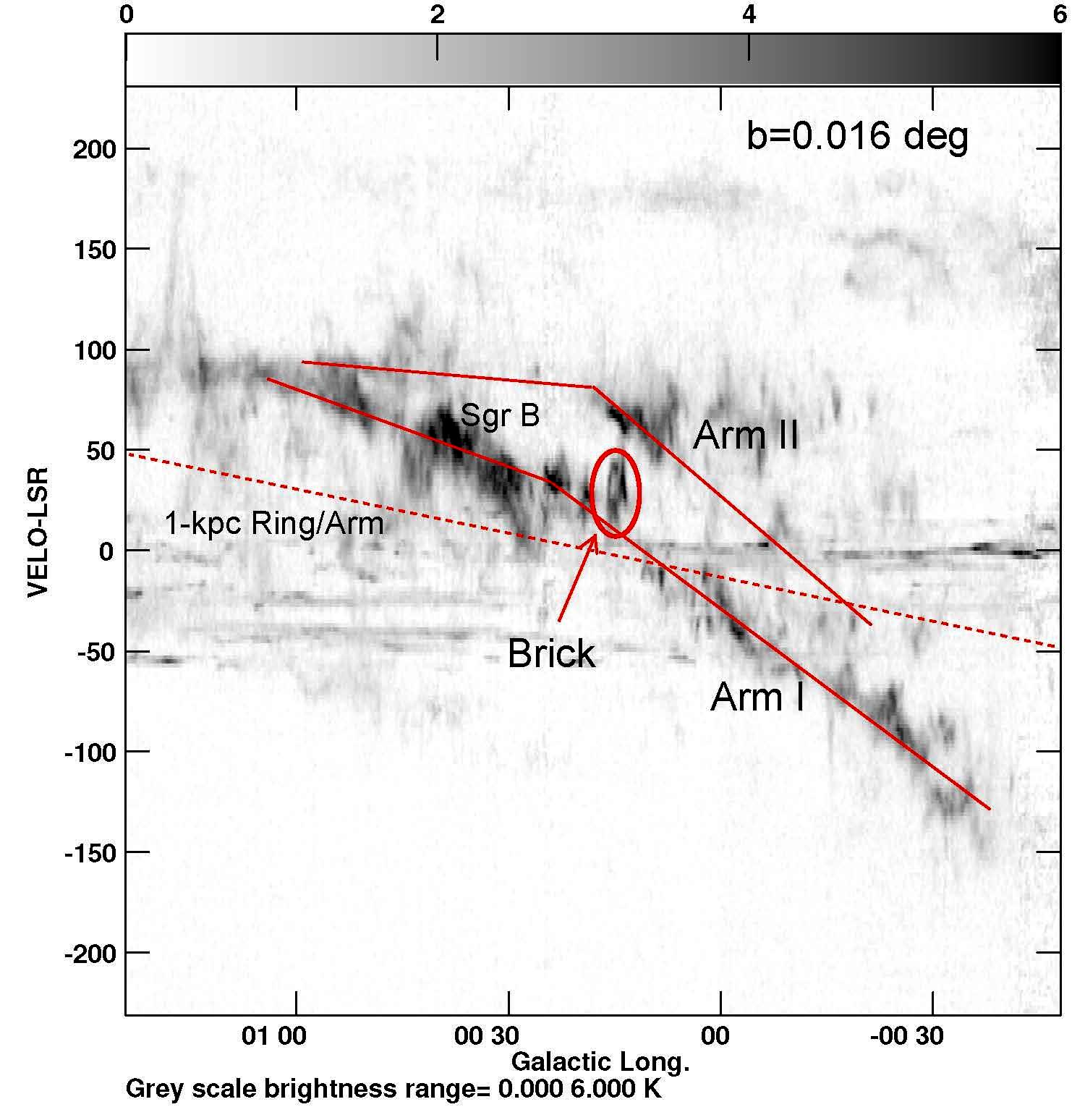}  \\
\vskip 3mm
\includegraphics[width=7.8cm]{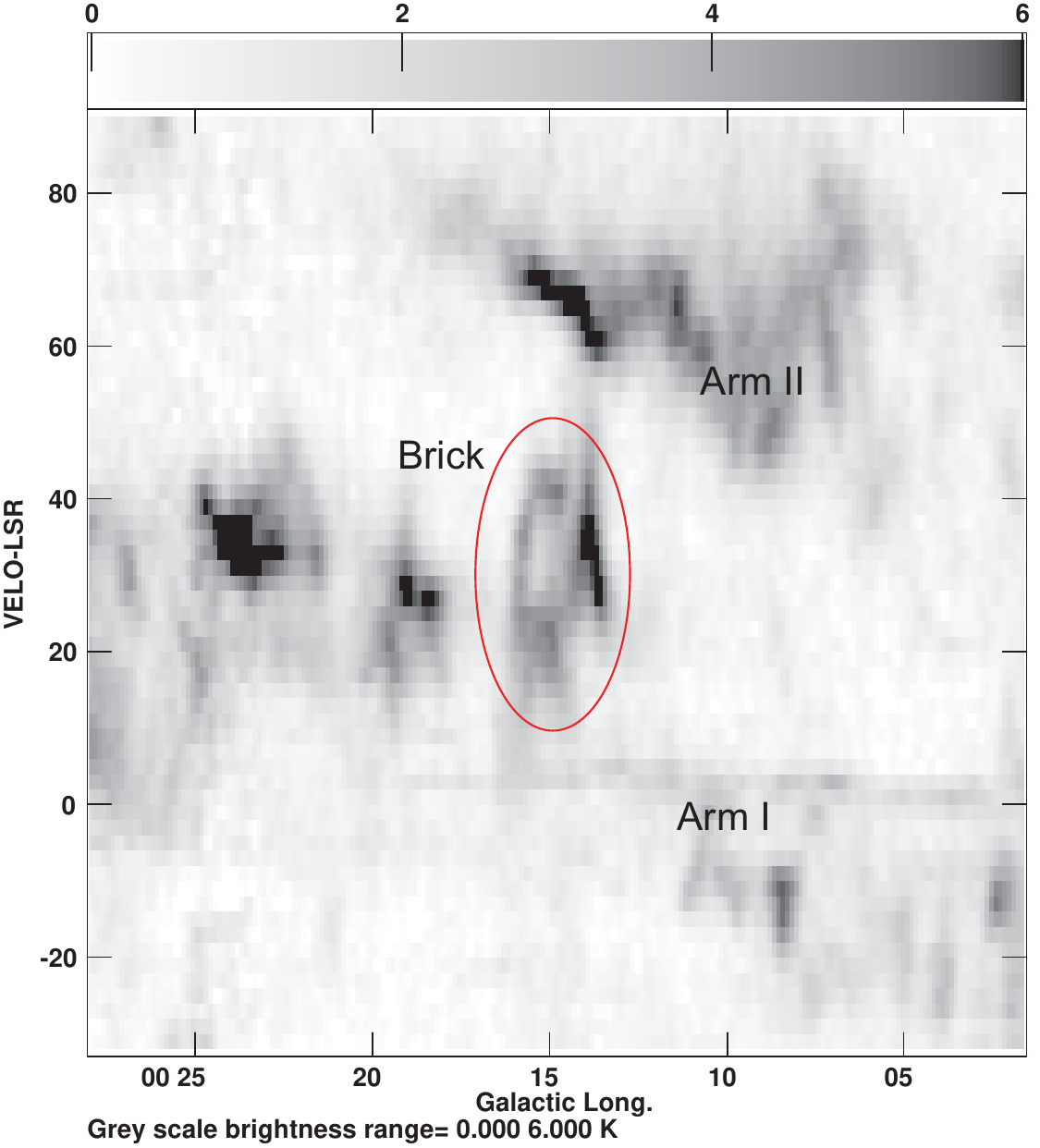}  
\end{center}
\caption{[Top] LVD of the \coth line across the center of Brick along $b=0\deg.016$. 
The Brick shows up as a clump at $l=0\deg.253$ and $\vlsr\simeq 30$ \kms between the GC Arm I and II (red lines) \citep{sof95}. Dashed line indicates a possible arm of galacto-centric radius 1 kpc on which the Brick might be located, if the distance is $\sim 7$ kpc, while such an arm is not recognized in this diagram. [Bottom] Same, but close up.
\red{Note: In both panels, the horizontal-coordinate unit is $dd\deg \ mm'$.
}}
\label{fig-lv-wide}	
\end{figure}

Figure \ref{fig-lv-wide} shows LVDs in wider area across G0.253+0.016 at a constant latitude $b=0\deg.016$, covering the central molecular zone (CMZ) (top panel) and close up (bottom).
The prominent ridge running from top left to bottom right represents the Galactic Center Arm I and the fainter one at higher velocities is Arm II \citep{sof95}.
The Brick is recognized as a clump at a systemic velocity  of $\vlsr = 30$ \kms ranging from $\sim 20$ to 40 \kms.    
The Brick is located in touch but slightly displaced from GC Arm I.
This will be used later for discussing the distance of the Brick, while kinematic distance using the Galactic rotation is not applied here because of the nearly zero longitude,

\section{Distance and kimenatics}

\subsection{Distance and location in the CMZ}

Extinction study of the bulge stars in infrared indicated a distance of $d= 7.2\pm 0.2$ kpc, locating the cloud in front of the central bulge \citep{zoc21}.
On the other hand, color-magnitude study of bulge stars toward the Brick \citep{nog21} and extinction of stars with known proper motion \citep{mar22} indicated that the Brick is located inside the Galactic bulge, and hence is likely a cloud located inside the CMZ.

If $d\sim 7$ kpc, the Galacto-centric distance of the Brick is $R\sim 1.0$ kpc. 
In order to see if the 1 kpc region is rich enough in the molecular gas for nesting such a dense cloud as the Brick, we examined an LVD from the Columbia 1.2-m CO-line survey \citep{dam01} in the central region at $l\sim \pm 7\deg$ ($R=R_0\sin l \sim 1$ kpc). 
We immediately found that the ring region at $R\sim 1 $ kpc is almost empty in the LVD.
Next, we examined the LVD of the CMZ in figure \ref{fig-lv-wide}, where we do not also find any LV feature parallel to the straight dashed line with gradient $dv/dl \sim 30$ \kms degree$^{-1}$ representing a supposed 1-kpc ring or arm rotating at $\sim 200$ \kms.
Therefore, it is difficult to attribute the Brick to a molecular disc or ring in the Galactic plane with radius $R\sim 1$ kpc, unless a completely isolated dense cloud is orbiting alone in such an empty region and is by chance observed exactly in the GC direction.

On other hand, the LVD in figure \ref{fig-lv-wide} suggests that the Brick is more closely correlated with the CMZ, which is composed of dense and clumpy molecular arms (GC Arms I and II) running at steeper slopes of $dv/dl\sim 150$ \kms degree$^{-1}$.
It is stressed that the systemic velocity of the Brick ($\sim 30$ \kms) is different only by $+10$ \kms from the ridge velocity at $\vlsr=+20$ \kms of Arm I at $l\sim 0\deg.25$, safely within the velocity dispersion of the CMZ molecular gas.
In fact, Arm I is full of clumpy clouds whose velocity displacements are $\sim \pm 20$ \kms with the extreme case of $\sim -30 \ekms$ of the Sgr B molecular complex.

It seems, therefore, more reasonable to consider that the Brick is an object physically associated with the CMZ rather than to locate it at 1 kpc away from the GC.
We, therefore, assume that the Brick is located inside the CMZ associated with GC Arm I.
Because Arm I composes a ring of radius $R\sim 160$ pc \citep{sof22}, the distance to the Brick is  $d=8$ kpc from the Sun for the GC (Sgr A$^*$) distance $R_0=8.2$ kpc.
This assumption is in agreement with the color-magnitude and extinction studies of stars of GC stars  that locate the Brick inside the CMZ \citep{nog21,mar22}.  

\subsection{On the heavy extinction}

We comment on the heavier extinction of the Brick than that of Sgr B molecular complex despite comparable or higher gas density in the latter.
Since the Brick is at $0\deg.25$ from the GC, it is a silhouette against the central core of the stellar bulge near $R\sim 35$ pc.
On the other hand, Sgr B at $l\sim 0\deg.6$ is silhouetted against the bulge stars at $R\sim 86$ pc.
It is known that the central stellar mass distribution is expressed by two components: one is the central bulge with scale radius 120 pc and central stellar density $2\times 10^2 \Msun$ pc$^{-3}$, and the other is the inner bulge or core which has the scale radius of 38 pc and center density $4\times 10^4 \Msun$ pc$^{-3}$ \citep{sof13}.
The difference of the distances of the clouds from the GC on the sky, 35 and 86 pc, therefore, yields a significant difference of the background infrared brightness with or without the inner bulge.
This results in much brighter background toward the Brick $\gtrsim 10^2$ times than for Sgr B, and explains the particularly heavy extinction measured in absolute brightness toward the Brick.
 
\subsection{Size of the Brick}

\begin{figure}   
\begin{center}   
\includegraphics[width=8cm]{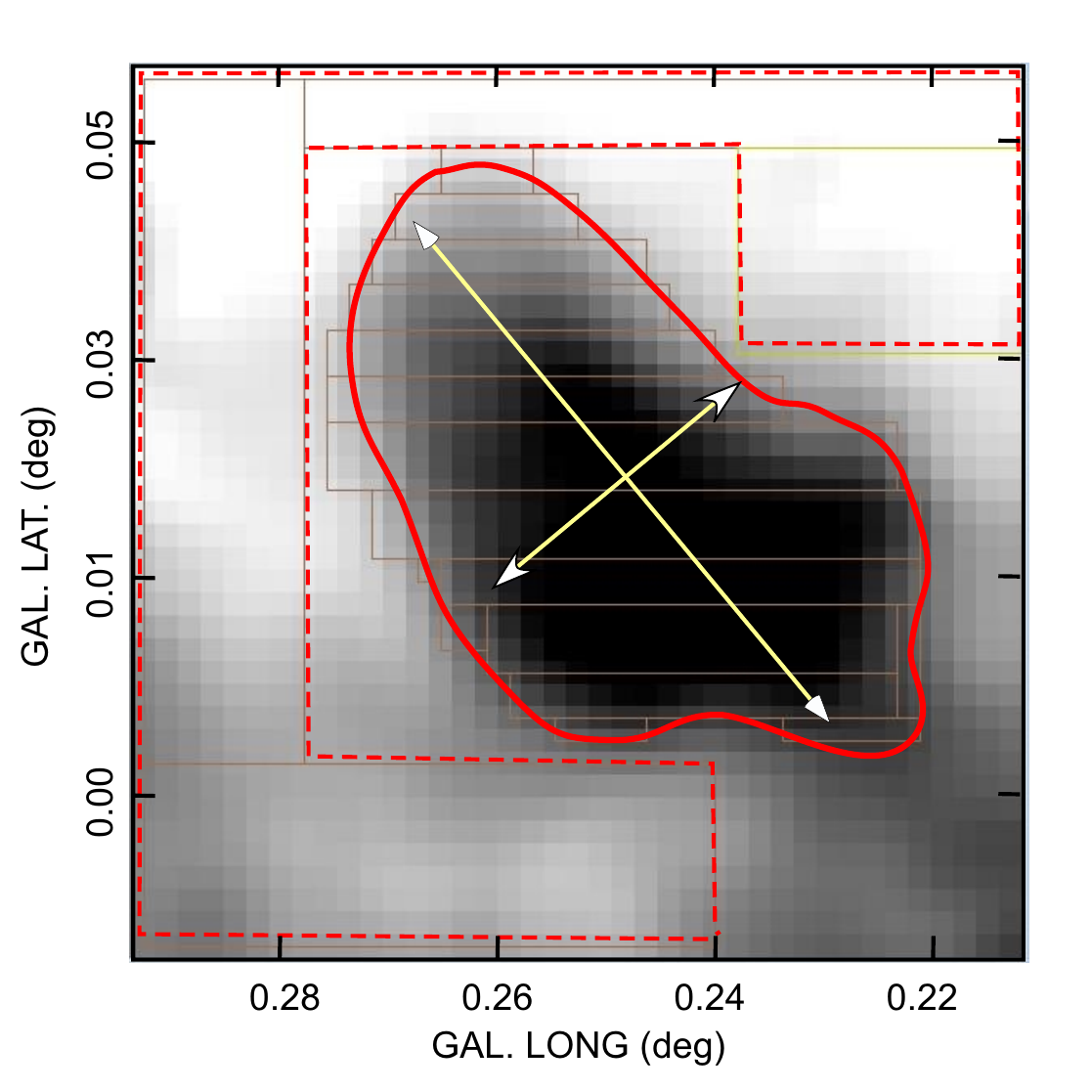} 
\end{center}
\caption{\red{\co moment 0 map of the Brick. The arrows indicate major and minor-axial diameters ($D_x,D_y$) as read on the \coth-line moment 0 map. Red line encloses the area for the \co-luminosity measurement in order to determine the molecular mass using the conversion factor. The dashed line encloses the area used to calculate the base level.}
}
\label{fig-onoff}	
\end{figure}

We use the moment maps to measure the fundamental parameters for calculating the kinematical parameters such as the size, molecular mass, kinematic and gravitational energies, the density, and time scale of the cloud.
\red{Figure \ref{fig-onoff} shows the \co-line moment 0 map, where the measured sizes in the \coth map is shown by the arrows, and the area for luminosity measurement by red line and off-source regions by dashed line.}

The derived parameters are listed in table \ref{tab}. 
FWHM  (full width of half maximum) sizes (diameters) $D_x=0\deg.060$ and $D_y=0\deg.030$ in the major and minor axial directions, respectively, of the Brick were measured by reading the coordinates at the steepest sides of the \coth-line profiles, which have both plateaued shapes. 
We thus obtain
$D_x=8.43$ and $D_y=4.13$ pc, respectively, for $d=8$ kpc,
and the size radius 
\be
r=\sqrt{D_x D_y}/2=2.96 \ \epc.
\ee 
\red{This size is slightly larger than that measured on the dust emission map of 2.7 pc for a distance of 8 kpc (originally 2.8 pc for a distance of 8.4 kpc) \citep{lon12}.}

\subsection{Molecular mass by \co-to-\Htwo conversion}
\label{sec_massenergy}
 
We first calculate the molecular mass using the conversion factor derived for the GC region of $\Xco=1.0\times 10^{20}$ \Htwo cm$^{-3}$ [\Kkms]$^{-1}$ \citep{ari96} and the nominal molecular weight of $\mu=1.38$. 
Using \co moment 0 map shown in figure \ref{fig-onoff} we measured the surface-integrated line intensity of the area enclosed by the red line.
We also measured the off-source mean intensity in the area enclosed by the dashed line.
By subtracting the corresponding surface-integrated intensity of the base level from the on-source value, we obtain the \co luminosity of the Brick,
\be 
L_{\rm 12CO}=2.21\times 10^{41}  {\rm K\ km\ s^{-1} cm}^2.
\ee
The total molecular-gas mass is then obtained to be
\be
\Mbrixco= \Xco L_{\rm 12CO} (2\mu \mH)=5.1\times 10^4\Msun.
\ee
The mean molecular-gas density is calculated by 
\be
\rho=\Mbrixco/(4 \pi r^3/3)=3.3\times 10^{-20}{\rm g\ cm}^{-3},
\ee
or 
\be
n_{\rm H_2}=6.5\times 10^3$ \Htwo cm$^{-3}.
\ee
The Jeans (free fall) time in the cloud is estimated to be
\be
t_{\rm J}=1/\sqrt{4 \pi G \rho}\sim 0.19\ {\rm  My}
\ee
for zero sound velocity limit. 
We also measured the peak intensity to be $\Ico=660$ Kkms, and the \Htwo column density at the brightest clump to be  
\red{$N_{\rm H_2}=\Xco \Ico=6.6\times 10^{22}$ \Htwo cm$^{-2}$. }

\subsection{The Virial mass}
  
The dynamical quantities can be estimated from the obtained CO kinematic quantities.
\red{Using the moment 2 map we measure the velocity dispersion to be $\sigma_{\rm cen}=13 \ekms$ near the cloud center and $\sigma_{\rm v}=10\ \ekms$ over the entire cloud. 
Since the central high dispersion may be influenced by the expanding bubble, as discussed later, we here adopt the overall value for the dispersion of the cloud. 
We here use mean velocity dispersion to calculated dynamical (Virical) mass as  
\be
\Mbrivir= r \sigma_v ^2/G \sim 6.8\times 10^4 \Msun.
\ee    
The density for this Virial mass is then calculated by 
\be
\rho=M_{\rm vir}/(4 \pi r^3/3)=
4.3\times 10^{-20}{\rm g\ cm}^{-3},
\ee
or 
\be
n_{\rm H_2}=8.7\times 10^3$ \Htwo cm$^{-3}.
\ee
The free-fall (Jeans) time of the cloud center is estimated to be
\be
t_{\rm J}=1/\sqrt{4 \pi G \rho}\sim0.17\ {\rm My}.
\ee
}  
Thus obtained total flux of the \co intensity was converted to the luminosity at a distance of 8 kpc, and converted to the molecular-gas mass by
\be M_{\rm brick}=\Xco \Ico A\times (2 \mH \mu) \ee
where A is the area of the Brick and $\mu=1.38$ is the mean atomic weight.
Adopting $\Xco=1.0\times 10^{20}$ \xcounit, we obtain 
$\Mbrivir=5.1\times 10^4 \Msun$.

The kinetic energy of the Brick corresponding to this Virial mass is $E_{\rm k}=1/2 M_{\rm vir} \sigma_v^2 \sim 0.68\times 10^{50}$ erg. 
The gravitational energy is estimated by
$E_{\rm g}=G M_{\rm mol}^2/r \sim 1.37\times 10^{50}$ erg,
trivially satisfying  
$2E_{\rm k}-E_{\rm g}\sim 0$.

\begin{table*}
\caption{Kinematic properties of the Brick G0.253+0.016 and the bubble.    }
\begin{tabular}{lll}    
\hline  \hline 
Parameter & Result &Remark\\
\hline 
\hline
{\bf Brick: observed quantities}\\
Distance $d=R_0-0.2$ kpc &8 kpc&On GC Arm I in front of Sgr A$^*$\\ 
Approximate centre $(l,b,\vlsr)$& $(0\deg.249,\ 0\deg.020, \ +30 \ekms)$ &Figure \ref{fig-onoff}\\
Major diameter (PA=$40\deg$) $D_x$& $0\deg.060=8.43 \epc$ &\coth mom. 0 map, figure \ref{fig-onoff}\\
Minor diameter (PA=$130\deg$)$D_y$&$0\deg.030=4.12 \epc$ & ibid\\
Size radius $r=\sqrt{D_x D_y}/2$  & 2.96 pc  \\ 
Velocity dispersion $\sigma_v$  &   10 \kms \\ 
\co Peak intensity $I_{\rm ^{12}CO}$ &  660 \Kkms   \\ 
\co Mean intensity $I_{\rm ^{12}CO}$ &  430 \Kkms  &Figure \ref{fig-onoff} \\ 

\hline
{\bf $\Xco$ mass}\\
Mass mol. $\Mbrixco$  & $5.1\xfour\Msun$ &\co m0 aperture photo.\\
Conversion factor $\Xco^{\rm GC}$&$1.0\times 10^{20}$ \xcounit &GC conv. factor \citep{ari96}\\ 
$\rho_0=\Mbrixco/(\frac{4\pi}{3} r^3) $ 
&$3.3\xmtwe$ g cm$^{-3}${\tiny $(=7.2\times 10^3$ \Htwo cm$^{-3}$)}
&For $\Xco^{\rm GC}$ mass \\ 
Energy, kinetic: $E_k=\Mbrixco\sigma_v^2/2$ &$0.51\times 10^{50}$ erg\\
Energy, gravi: $E_g=G \Mbrixco^2/r$ &$0.76\times 10^{50}$ erg \\ 
$t_{\rm ff}=1/\sqrt{4\pi G\rho}$  &0.19 My  \\ 

\hline
{\bf Virial mass}\\
Mass Virial: $\Mbrivir=r\sigma_v^2/G$ &$6.8\xfour\Msun$  & Virial mass\\  
New conversion factor $\Xbri$ & $1.3\times 10^{20}$ \xcounit & $\Xbri$ mass = Virial mass\\
$\rho_0=\Mbrivir/(\frac{4\pi}{3} r^3$) 
&$4.3\xmtwe$ g cm$^{-3}$ \tiny{($=9.3\times 10^3$ \Htwo cm$^{-3}$)}&ibid \\   
Energy kinetic: $E_{\rm k}=\Mbrivir\sigma_v^2/2$  &$0.68\xfifty$ erg& ibid \\ 
Energy, gravi. $E_{\rm g}=G \Mbrivir^2/r$  & $1.37\xfifty$ erg&$2E_{\rm k}-E_{\rm g}=0$ (Vrialized)\\ 
 $t_{\rm ff}=1/\sqrt{4\pi G\rho}$  &0.17 My  \\ 
\hline \hline
{\bf Bubble: Buried SNR}\\ 
Centre position $(l,b,\vlsr)$&$(0\deg.245, \ 0\deg.018,\ +30 \ekms)$&Figure \ref{fig-bub}\\
Radius $\rbub$  &$0\deg.0133=1.85$ pc&\\%
Mass $\Mbub= \frac{4\pi \rbub^3}{3} \rho_0$ 
& $1.7\xfour\Msun $ & $\rho_0$ from virial mass for stable Brick\\
Expansion velo $\vexpa$ & 10 \kms\\ 
Energy kin. $E_{\rm bub}=(1/2) \Mbub \vexpa^2$  
&$0.17\xfifty$ erg\\
Age (Sedov time) $t_{\rm sed}=(2/5)\rbub/\vexpa$  
&0.072 My \\
\hline 
\end{tabular}  \\ 
    \label{tab}
\end{table*}

\subsection{Comment on the conversion factor and gas-to-dust mass ratio}
\label{sec_xco}

The Virial mass is therefore 1.6 times greater than the molecular mass for the conversion factor in the GC of $Xco^{\rm GC}=1.0\times 10^{20}$ \xcounit.
This means either that the cloud is not Virialized, or that the conversion factor is wrong.
If the former is the case and the cloud is unstable and being disrupted, its age must be as short as $t\sim r/v\sim 3\times 10^5$ y. 
If the latter is the case, the mass corresponds to a larger conversion factor of $\Xco=1.6\times 10^{20}$ \xcounit, which is closer to the local value \citep{bol13}.
 
\red{In either estimates, the mass ($\sim 5-6.8\times 10^4 \Msun$, table \ref{tab}) of the Brick derived hear from the CO-line measurement is a factor of three smaller than the current measurement from the dust emission using a gas-to-dust mass ratio of 100, $M_{\rm bub;dust}\sim 1.3\times 10^5 \Msun$ \citep{lis91,lis94a,lon12}, while the scale radius hear (CO, 2.96 pc) is about the same as current measurements (dust, 2.8 pc).
The discrepancy may be solved if the gas-to-dust ratio is reduced to one third, or gas-to-dust ratio in the GC is $\sim 30$.
This would not be unrealistic because of the higher metallicity in the GC, as in the case of $\Xco$ \citep{ari96}.}

\subsection{Rotation}
\label{sec-rotation}
 
The moment 1 map (velocity field) in figure \ref{fig-mom} shows a clear velocity gradient along the major axis ($x$ axis) of the Brick at position angle $\sim 40\deg$ at $dv/dx\sim 2.5$ \kms pc$^{-1}$.
This gradient is consistent with that discussed by \citet{hen19}, while an order of magnitude smaller than that reported by \citet{hig14}.
If the velocity gradient is attributed to the rotation of the Brick along the minor axis, the rotation velocity  of the major-axis ends at $r\sim \pm 4 \ (=D_x/2)$  pc is $V_{\rm rot}\sim 6.5$ \kms. 
The centrifugal force due to this rotation is sufficiently smaller than the gravitational force by the Brick with a molecular mass of 5.1--6.8$\times 10^4\Msun$ (table \ref{tab}).  
 
\subsection{Radio continuum properties}

\begin{figure*} 
\begin{center}  
(a)~~~ \includegraphics[width=6cm]{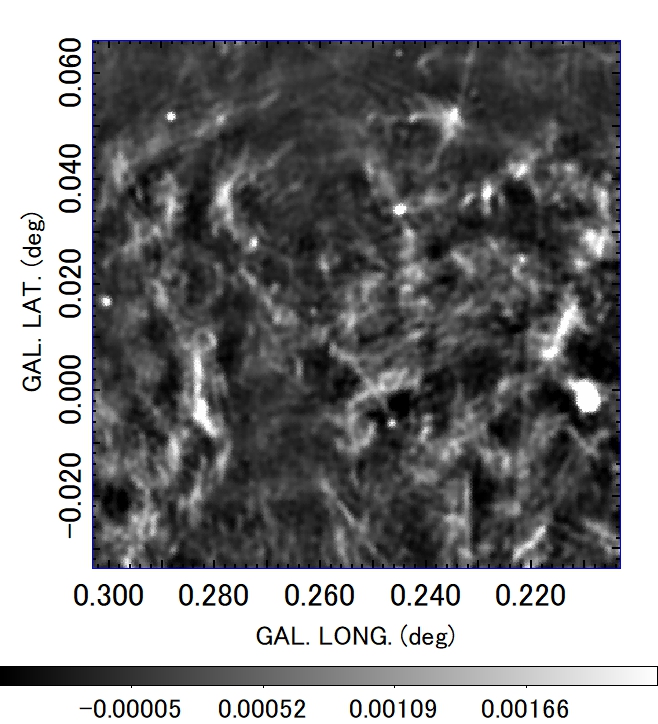}   \hskip 5mm
(b)~~~\includegraphics[width=6cm]{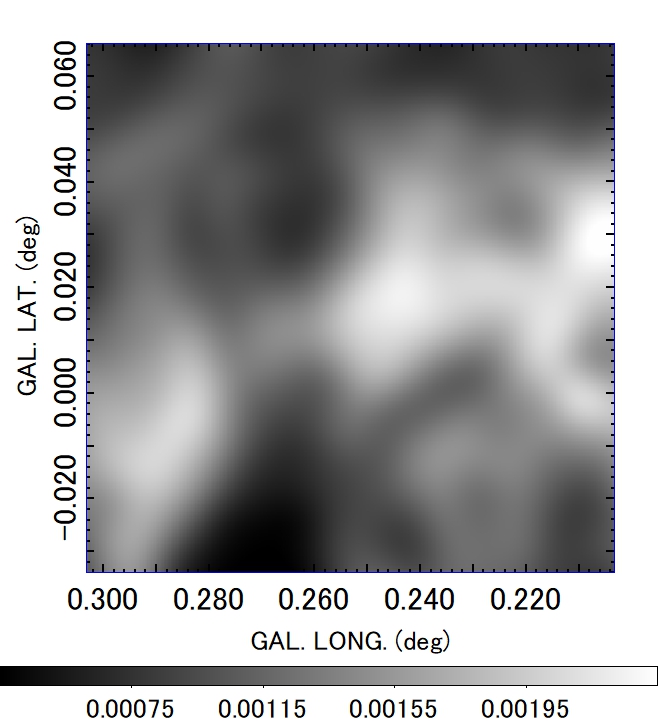}  \\ 
(c)~~~\includegraphics[width=6cm]{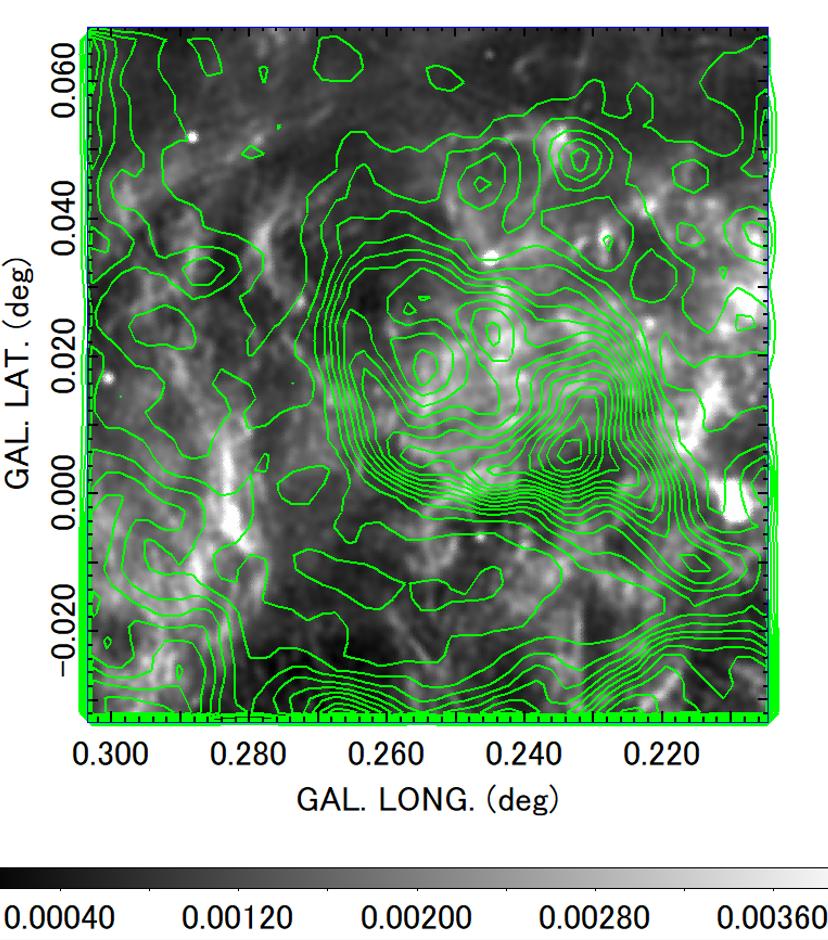}    \hskip 3mm
(d)~~~\includegraphics[width=6.2cm]{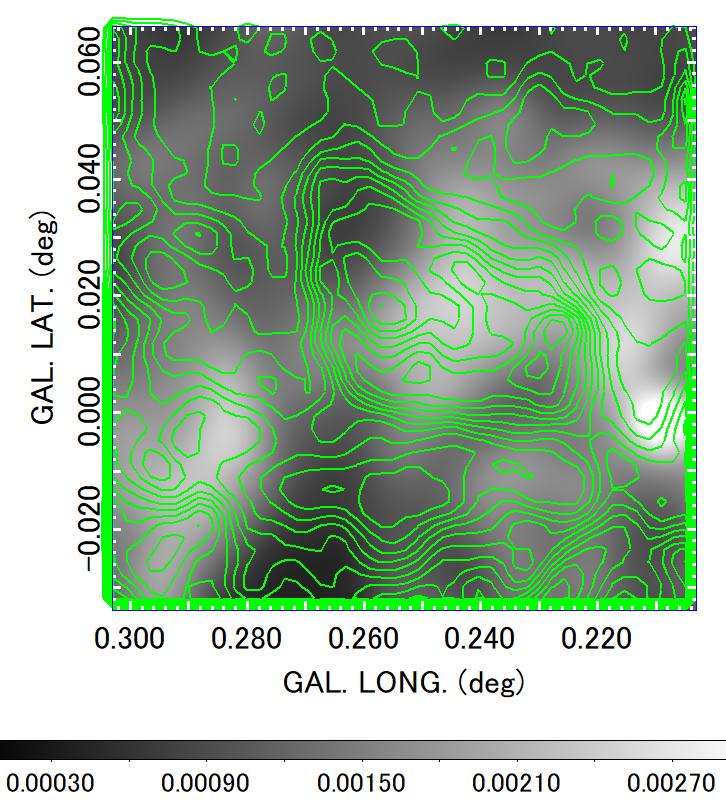}\\   
(e)~~~\includegraphics[width=6cm]{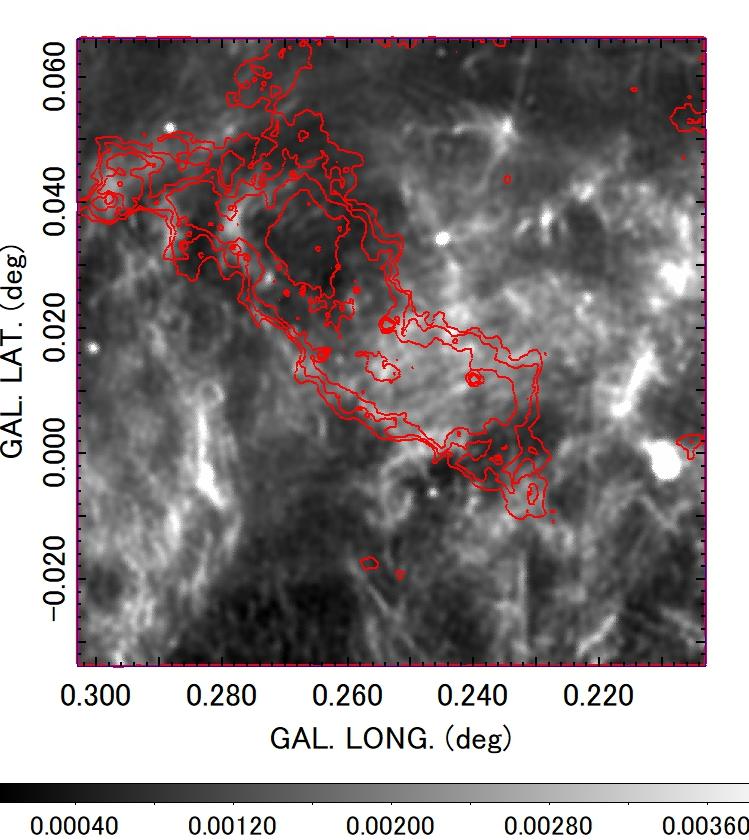}  \hskip 3mm
(f)~~~\includegraphics[width=6cm]{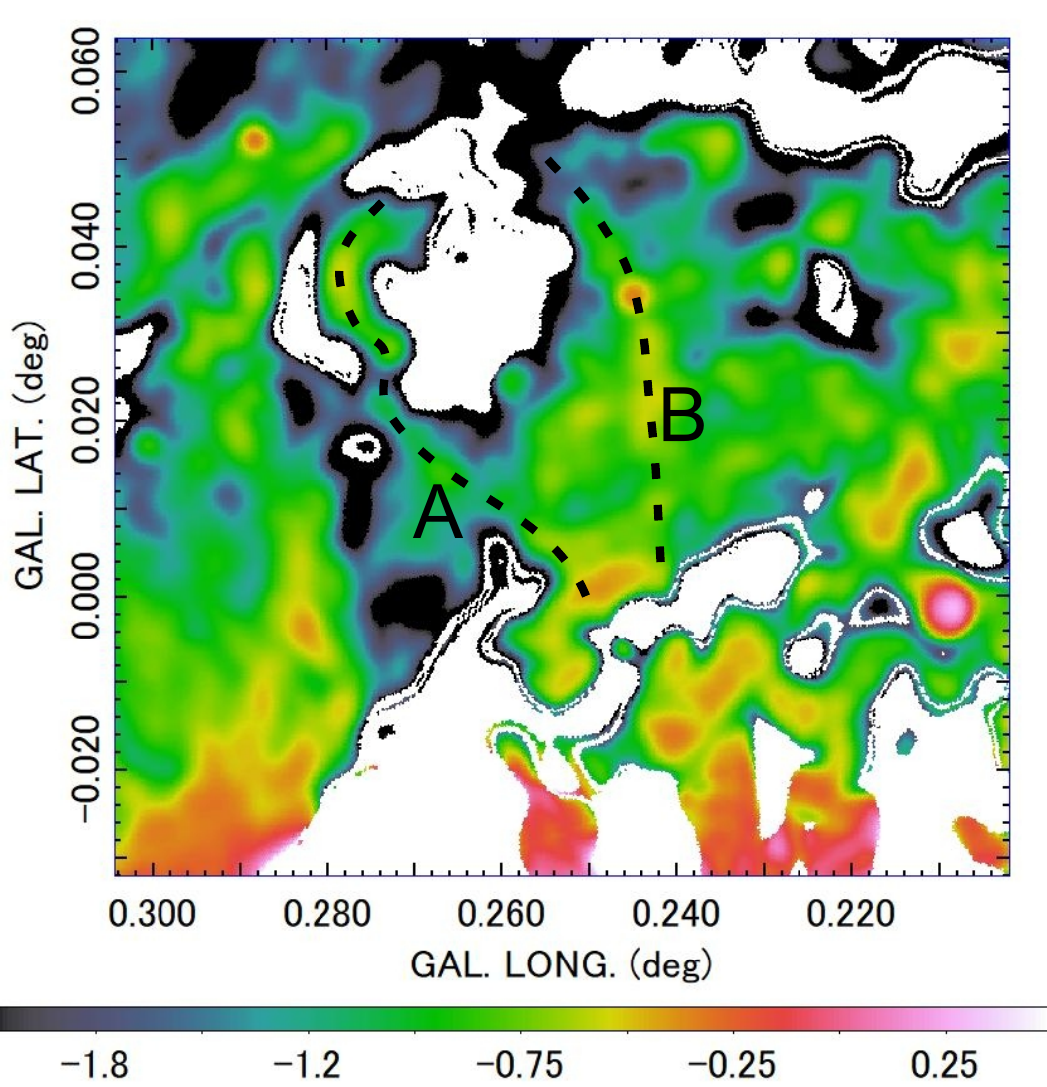}   
\end{center}
\caption{
\revtwo{
(a) 1.3 GHz radio continuum map from MeerKAT (grey scale by bar in Jy/4$''$ beam) \citep{hey22,yus22}, where the background emission with scale sizes greater than 15$''$ has been removed using the BGF method \citep{sof79}.
(b) Extended components with scale sizes greater than 15$''$. 
(c) Original MeerKAT 1.3 GHz image 
superposed by \coth\ $\Tb$ contours every 0.5K at $\vlsr=31$ \kms. Note: Map a + Map b = Map c.
(d) Smoothed 1.3 GHz map simply convolved with a Gaussian beam of FWHM $15''$ superposed by \coth contours at 29 \kms.
(e) Same as (c), but overlaid by mid-infrared intensity by 8 $\mu$m contours from 100 (minimum) to 300 every 100 mJy/sr.  
(f) Distribution of the spectral index near 1.3 GHz, showing that the radio continuum emission  is mostly nonthermal. 
Dashed lines A and B trace the radio filaments.}
}
\label{fig-mkat}	
\end{figure*}

\revtwo{Figure \ref{fig-mkat} shows  radio-continuum surface brightness map at 1.3 GHz extracted from the GC survey with the MeerKAT \citep{hey22} at resolution $4''$. 
Panel (a) shows a background-filtered map of the MeerKAT image using the BGF method \citep{sof79}, where map (b), showing the extended components with scale sizes greater than $15''$, has been subtracted. 
Panels (c) and (e) show overlays of the \coth-line $\Tb$ at $\vlsr \sim 30$ \kms by  contours every 0.5 K, and 8 $\mu$m intensity also by contours from the Spitzer survey \citep{chu09}, respectively. 
Panel (d) shows the same, but CO contours are overlaid on a simply smoothed 1.3 GHz map. 
Panel (f) shows the distribution of radio spectral index near 1.3 GHz, indicating the radio emission is mostly non-thermal with $\alpha \lesssim -0.5$ ($\Sigma_{\rm radio}\propto \nu^\alpha$) \citep{hey22,yus22}. 
No signature of thermal emission with flat spectrum from HII regions is recognized here.
The dashed line A and B in panel (f) represents the radio filaments that show up in the upper panels.}

Two horn-like filaments (A) and (B) parallel to the eastern and western edges, respectively, of the Brick compose a lobe structure at position angle $\sim 30\deg$.
Filament A apparently coincides with the eastern dust \citep{joh14} and molecular arcs \citep{hig14,hen22}.
The western filament (B) and enhanced emission near the bubble center as observed at 5 GHz with the VLA (Very Large Array) \citep{hen22} are also recognized in this map.

As to physical association of the radio features with the Brick, the following two cases are considered:
i) They are part of an SNR originating in the Brick.
ii) They are background GC filaments not related to the Brick. 

If (i) is the case, we may estimate the diameter using the surface brightness-diameter ($\Sigma-D$) relation for SNRs \citep{cas98}.
$\Sigma_{\rm 1.3\ GHz}$ of the filaments is read from the radio image to be $\sim 2$ mJy beam$^{-1}$ for $3.6''$ synthesized beam.
Assuming filament's coverage over the supposed SNR coinciding with the molecular bubble to be on the order of $\sim 0.1$ and spectral index $-0.7$, we obtain a rough estimate of radio surface brightness at 1 GHz:
$\Sigma_{\rm 1\ GHz}\sim 6.6\times 10^{-21}$ W m$^{-2}$ Hz$^{-1}$ sr$^{-1}$.
Applying the $\Sigma_{\rm 1 GHz}-D$ relation \citep{cas98}, we obtain $D\sim 30$ pc.
Combining this diameter with the angular size of $\sim 1'6$, the distance is estimated to be $\sim 65$ kpc, which obviously contradicts the distance to the Brick near the GC.  

We may, therefore, conclude that case (ii) is more plausible.
The radio features are not likely to be physically associated with the Brick, but may be parts of the background emission in the direction of the Galactic Centre.
In fact, the GC is full of  non-thermal filaments \citep{hey22,yus22,sof23}.
\revtwo{In addition, this direction is contaminated by numerous radio sources along the line of sight over $\sim 20$ kpc through the Galactic disc crossing many spiral arms. } 
 
From the radio map, we can also estimate an upper limit to the thermal emission from the Brick, assuming that the brightness of the thermal emission is less than the brightness fluctuation of the non-thermal emission on the order of $\lesssim 0.5$ mJy beam$^{-1}$ at 1.3 GHz. 
This yields an upper limit to  the emission measure as 
$EM\lesssim 1.6\times 10^4$ pc cm$^{-6}$, which is less than that for the weakest HII regions in the Galaxy \citep{dow80}.
This estimation is consistent with the negative reports of massive-star formation in the Brick \citep{lon12,joh14,hen22}.

\section{Dark SNR buried in the Brick}

We have shown that the Brick nests an expanding molecular bubble from the channel maps and position-velocity diagrams, while it is not associated with radio continuum emission indicative of any HII region or radio SNR. In this section, we derive the physical parameters of the bubble, and explain it as a dark SNR buried in the molecular Brick \citep{luc20,sof20,sof21}. 
We also compare our model with those in the literature.

\subsection{Expanding bubble}
\label{sec_bubble}

In figure \ref{fig-chan} we showed channel maps of $\Tb$ in the \coth line, which exhibited a bubble structure as marked by the red circle.
Besides the spatial bubble structure, the position-velocity diagrams in figures 
\ref{fig-chan-lvd} and 
\ref{fig-chan-bvd} showed elliptical features, which indicates that the bubble is expanding at 
$v_{\rm exp}\sim 10$ \kms.

\begin{figure*}  
\hskip 4cm {\Large $(l,b)$} \hskip 6cm {\Large $(\vlsr,b)$}\\
\includegraphics[height=8.5cm]{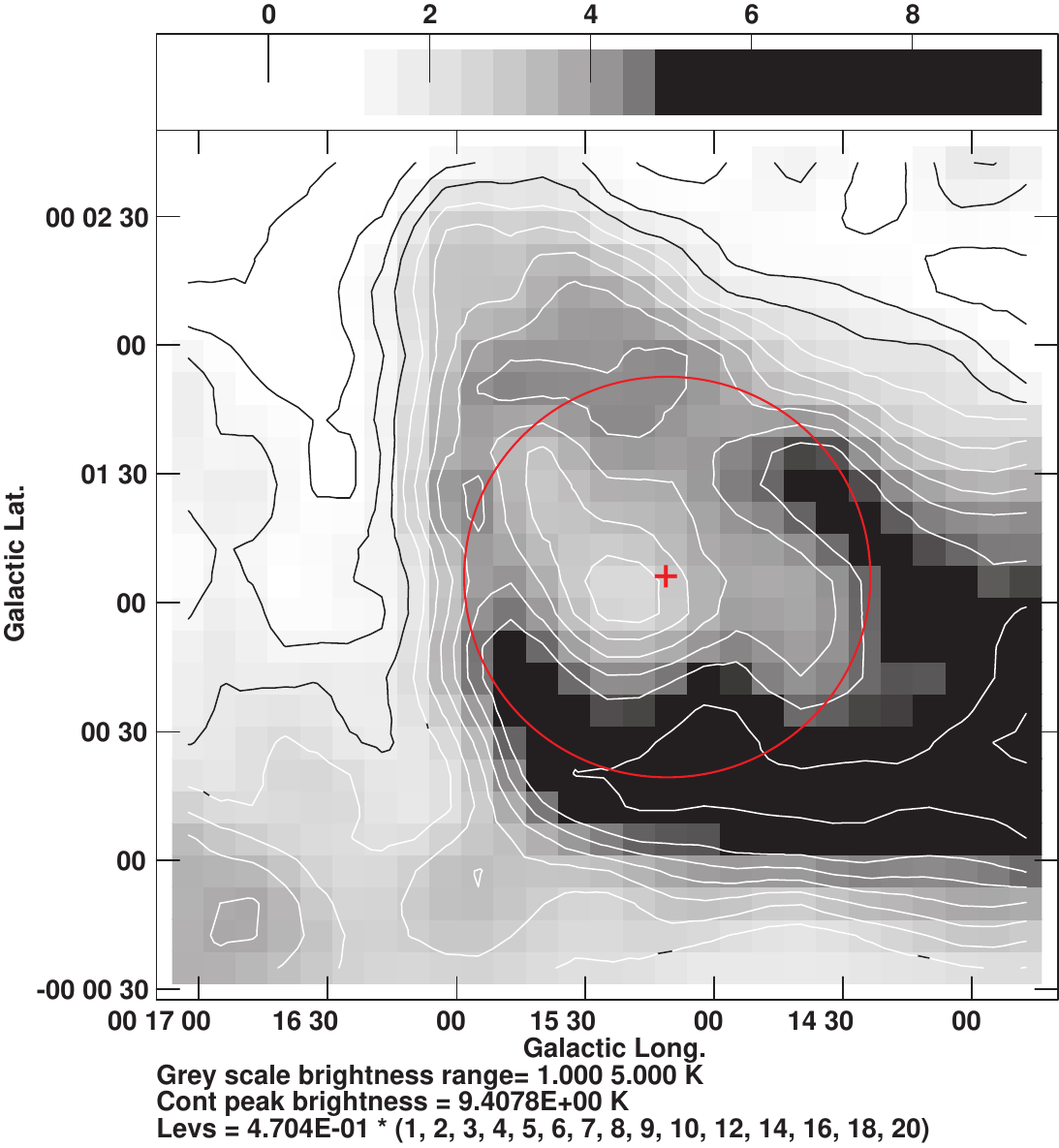}\\
\vskip -8cm 
\hskip 8.5cm \includegraphics[height=8.5cm,angle=90]{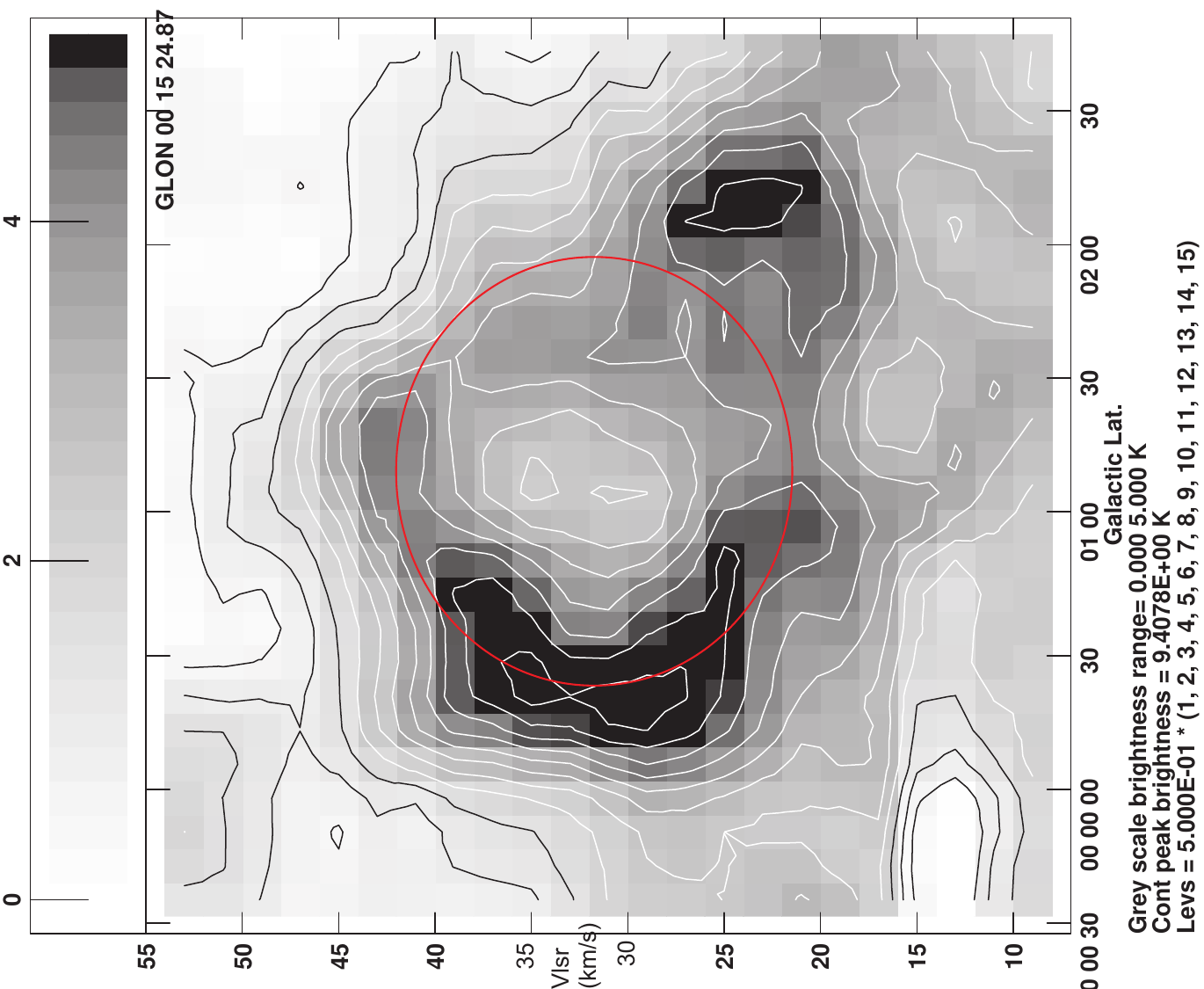}\\ 

\vskip 1cm 
{\Large \hskip 4cm $(l,\vlsr)$ \hskip 5cm $T_{\rm b}(\vlsr)$}\\

\hskip 1cm  \includegraphics[height=8.5cm]{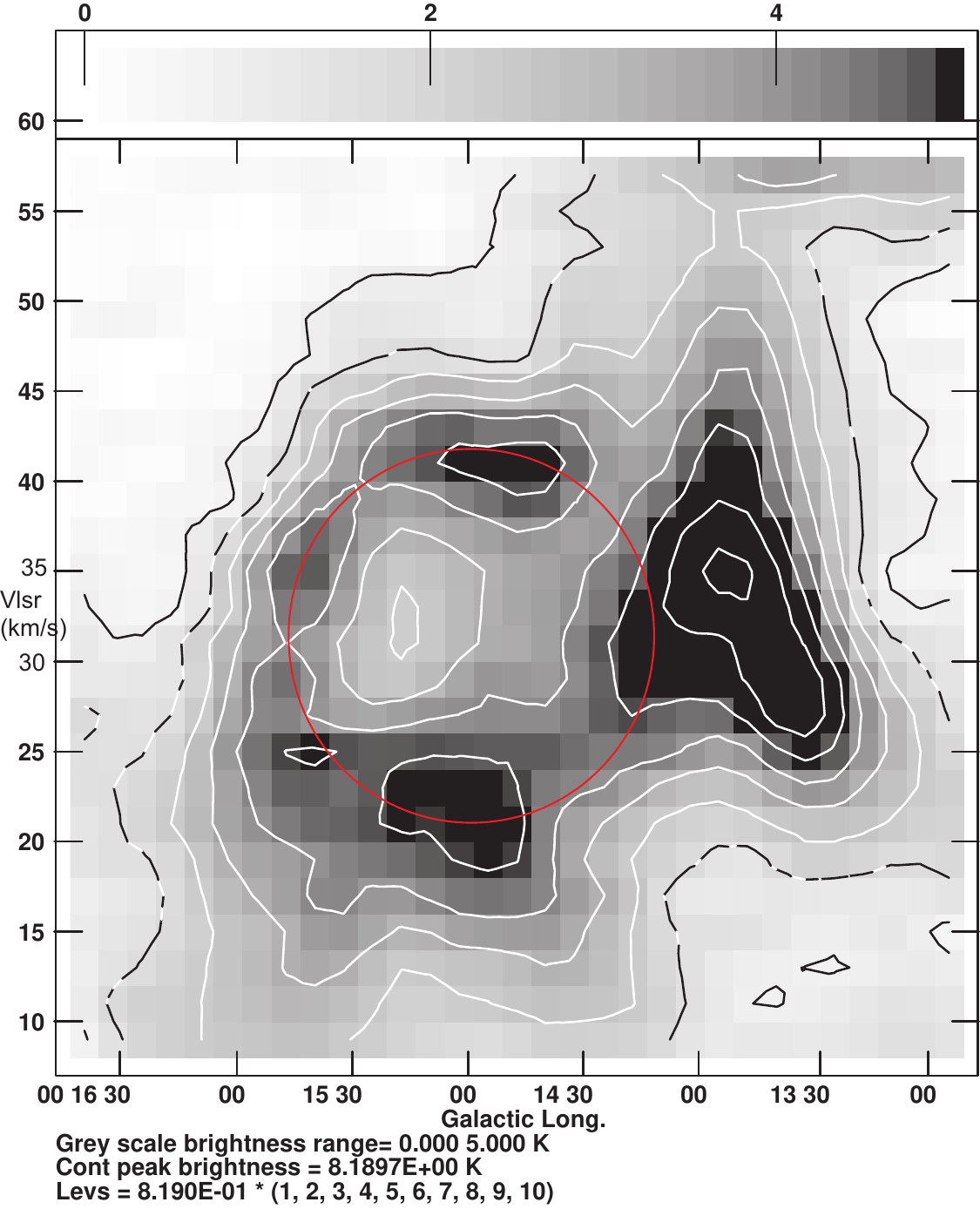} 
\hskip 5mm \includegraphics[height=8cm]{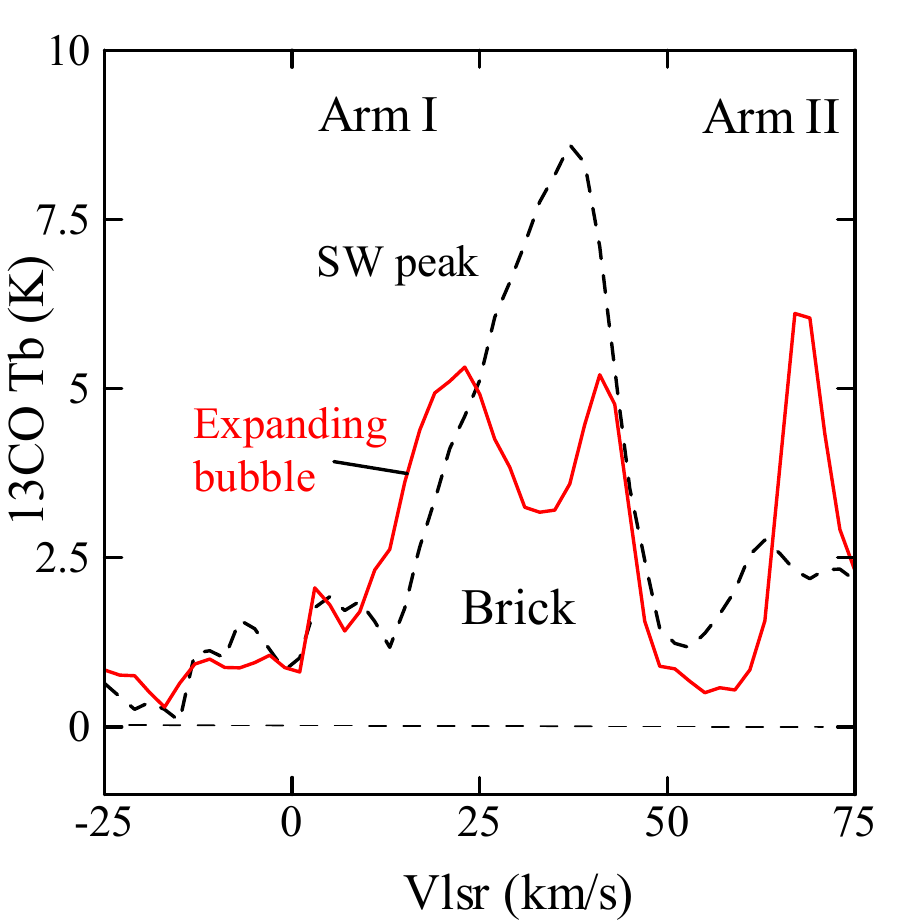}    
\vskip 5mm
\caption{
[Top] \coth $Tb$ map at $\vlsr=29$ \kms, showing a spherical shell structure. 
The circle approximately trace the bubble with radius $r_{\rm bub}=0\deg.0132=1.84 \epc$ centered on $(l,b)=(0\deg.245,0\deg.018)$ marked with the cross.
Bar and contours indicate $\Tb$ in K.
[Top right] $(b,\vlsr)$ diagram (vertical slice) across the bubble center.
Panel is rotated by $90\deg$.
[Bottom left] Same, but $(l,\vlsr)$ (horizontal slice). 
[Bottom right] \coth line spectra toward the bubble center (red), showing symmetric expansion in the line of sight. 
\revtwo{The double-peaked profile in the center of the shell is due to the expansion along the line of sight, which yields the maximum velocity dispersion near the center of moment-2 map in figure \ref{fig-mom}.}
Dashed line shows a peak in the SW corner of top-left panel. 
Note: Longitude and latitude coordinates are in unit of $dd\deg \ mm' \  ss''$.}
\label{fig-bub}	
\end{figure*}

Figure \ref{fig-bub} enlarges the channel and position-velocity maps at representative positions and velocities, as  marked by the red ellipses, which represent a $(l,b,\vlsr)$ sphere with radius 
$0\deg.0133=1.85$ pc (at 8 kpc distance) centered on 
$(l,b,\vlsr)=(0\deg.245,\ +0\deg.018, \ +30 \ekms)$.  
The bottom-right panel shows a \coth line spectrum toward the bubble center, showing a  double-peak profile with blue- and red-shifted peaks typical for expanding motion in the line of sight.

We stress that the expansion is nearly symmetric with respect to the  bubble's centre not only on the sky but also in the velocity directions.
This indicates that the bubble is a closed spherical structure without a break totally embedded in the Brick.  

The centre position of the CO bubble is slightly shifted to the south west by $\sim 0.6$ pc from the geometrical center of the Brick at $(l,b)\simeq (0\deg.253,+0\deg.016)$.
The south-eastern edge of the bubble coincides in position with the dust arc, whose center is at about the same position as the present CO-bubble's centre \citep{hig14,hen22}.

\subsection{Mass and energy of the expanding bubble}

\citet{hen22} have derived the mass of the arcs in the Brick, which coincide with the eastern limb of the present CO-line bubble, to be $M_{\rm arc}\sim 3\times 10^3\Msun$ and kinetic energy of the expanding motion of the arc $E_{\rm k: arc}\sim 7\times 10^{47}$ erg and momentum $p_{\rm arc}\sim 1.4\times 10^4 \Msun$ \kms. 

We here derive the mass and energy of the bubble from our CO-line data (figure \ref{fig-bub}).
However, it is ambiguous to abstract the mass properly belonging to the bubble from the intensity maps presented in this paper.
Therefore, we assume that the bubble is a structure whose mass has been plowed from the cavity, which had approximately the same density as the mean density determined in the previous subsections.

We assume that the Brick is gravitationally stable, and the total mass is equal to the Virial mass, which yielded  mean density of $\rho_0\sim 4.3\times 10^{-20}$ g cm$^{-3}$ (table \ref{tab}).
The total mass of the bubble is then estimated as
\be
\Mbub\sim \rho_0 (4\pi r^3/3)\sim 1.7\times 10^4 \Msun,
\ee 
and the kinetic energy of the expanding motion is
\be 
E_{\rm kin}\sim 1/2 \Mbub v_{\rm exp}^2 \sim 1.7\times 10^{49} {\rm erg}.
\ee 
The derived quantities are listed in table \ref{tab}.

We point out that the mass and energy of the expanding bubble are smaller than those of the whole Brick by a factor of 4.
This means that the bubble is not significantly disturbing the entire structure of the Brick. 

\subsection{Current models for the bubble}
\label{sec_models}

There have been two major ideas to explain the observed arc-shaped properties of the Brick based on interferometric observations of other molecular lines: 
The cloud-collision model \citep{hig14} and stellar-feedback model \citep{hen22}.
The latter may be categorized into cases that take account of expansion of an HII region and stellar wind from central early-type stars. 

\subsubsection{Cloud-collision model}

Cloud-collision model \citep{hig14} postulates collision of a compact cloud of mass $\sim 0.5 \times 10^5 \Msun$ and radius $\sim 1.5$ pc against a cloud of $\sim 2\times 10^5 \Msun$ and $\sim 3$ pc at velocity of $\sim 30$--60 \kms, where the masses are taken from the determination using the dust emission for gas-to-dust mass ratio 100 \citep{lon12}.
Difficulty in this model would be the absence of ionized dense gas (HII region) inevitably created by such high-speed, on-going collision.
Another concern is the long mean free path and collision time, which are calculated to be $L_{\rm col}\sim 10$ kpc and $t_{\rm col}=L_{\rm col}/v\sim 200$ Myr, if the CMZ is filled with similar-mass clouds in radius $\sim 200$ pc and full thickness $\sim 56$ pc with total molecular mass $2.3\times 10^7 \Msun$ \citep{sof22}.
This collision time is three orders of magnitudes longer than the Jeans time, $t_{\rm J}\sim 7\times 10^4$ y of the smaller cloud, which leaves a question how the cloud survived for such long time before collision.
One more concern is their orbits: 
why did the colliding cloud come from the halo direction at altitude angle $\sim 50\deg$, as the morphology of the arc indicates, and how was the angular momentum between the two unbound clouds removed in order to make the head-on collision?
Therefore, unless the orbits of the two clouds are determined by observations, we may consider other scenarios for the bubble formation in the Brick. 

\subsubsection{Stellar-wind model}

The wind-blown bubble model postulates a shell structure with molecular mass $\sim 3\times 10^3 \Msun$, kinetic energy $\sim 7\times 10^{47}$ erg, and momentum $\sim 1.4\times 10^4 \Msun \ekms$ from the interferometric observations with ALMA \citep{hen19,hen22}, and two possible scenarios have been proposed:
One is that the arc is formed by thermal pressure of HII gas ionized by the central OB cluster, which is however, ruled out in view of the insufficient amount of UV photons by the adopted model.
The other, which the authors prefer, is that the arc is driven by stellar winds from a cluster of $\sim 10^3 \Msun$ in the center, which is hidden behind the dusty cloud. 
However, the presently derived mass, energy and momentum are an order of magnitude greater than those used in the wind model, which might be difficult to be explained by the wind model. 
In the following subsection, we propose an alternate model, which assumes a supernova explosion in the centre of the Brick.

\subsection{Buried SNR model}
 
We try to explain the observed energy and morphology of the molecular expanding bubble by a buried supernova remnant (SNR) in the Brick.
The bubble structure can be approximately traced using the Sedov relation by assuming adiabatic expansion after a point explosion in the cloud:
\be 
E_0 \sim 1/2 M v^2,
\label{eqE}
\ee
 where 
$E_0$ is the input energy by the SN explosion,
$ 
M\sim4 \pi r^3\rho_0/3
$ is plowed gas on the shell, 
$v=dr/dt$ is the expansion velocity, and 
$\rho_0$ is the ambient density of the ISM.
The relation is equivalent to the Sedov's solution, and reduces to
\be 
v=dr/dt\sim a \ r^{-3/2}, 
\label{eqv}
\ee
and is solved to give the radius as a function of time,
\be 
r\sim  b \ t^{2/5}, 
\label{eqr}
\ee
and the age by the radius and velocity,
\be 
t\sim (2/5) r/v. 
\label{eqt}
\ee  
Here,  
$a=[2E_0/(4\pi \rho_0)]^{1/2}$ and 
$b=(5/2)^{2/5}a^{2/5}= 
1.256 (E_0/\rho_0)^{1/5}$ are constants. 
Inserting $v=10.0$ \kms, $r=1.85$ pc,  we obtain
$t\sim 7.2\times 10^{4} \ {\rm y}$.  
 
We recall that the gas density in the Brick is four orders of magnitudes higher than that in the interstellar space of the Galactic disc, where the majority of the known SNRs of shell type have been discovered. 
Equations (\ref{eqv}) and (\ref{eqr}) indicate that the SNR in the present circumstance evolves much more quickly than those in such circumstances, so that the emission phase was over in the early stage ($t\lesssim 10^{2-3}$ y) by exhausting the energy via strong thermal emission of ionized gas \citep{shu80,whe80,luc20}.
After the bubble cooled down and became radio quiet, the shell is still expanding due to the conservation of the energy and momentum.

Since the kinetic energy of the expanding motion is several times smaller than the gravitational energy of the Brick, the SN explosion does not disturb the stability of the entire cloud.
\revtwo{The passage of the shock wave would excite turbulence within the molecular shell, increasing the velocity dispersion to a value comparable to the expansion velocity.
This value is around $\sigma_{\rm turb}\sim v_{\rm expa}\sim 10$ \kms, comparable to the general dispersion in the entire brick.
However, it would be difficult to distinguish such a narrow velocity component, if any, from the general line profiles of the molecular cloud.
}
 
 
\section{Summary}
We obtained detailed kinematics and energetics of the GC Brick G0.253+0.016+30 \kms by analyzing CO line data obtained with the Nobeyama 45m mm-wave telescope at a resolution of $15''$.
Close inspection of the longitude-velocity diagram of the CO line emission shows that the Brick may be within the CMZ, but in front of the GC associated with GC Arm I.
This places the Brick at a distance of 8 kpc, about 0.2 kpc in front of the GC.

We showed that the Brick is a dense molecular cloud with Virial mass $\Mbrivir\sim 6.8\times 10^4 \Msun$ and gravitational energy $E_{\rm g}\sim 1.37\ \times  10^{50}$ erg and kinetic energy $E_{\rm k}\sim 0.68\times 10^{50}$ erg.
By adopting the Virial mass for the molecular gas mass, we obtain a new CO-to-\Htwo conversion factor of $\Xbri\sim 1.3\times 10^{20}$ \xcounit.
 
The center of the cloud is a cavity surrounded by a dense molecular bubble, which is expanding at $\vexpa=10$ \kms with kinetic energy $E_{\rm k, bub}\sim 1.7\times 10^{49}$.
However, the bubble does not nest radio continuum SNR or HII regions, and is therefore not related to massive star formation.
We suggest that this bubble could be a dark SNR (dSNR) buried in the Brick, whose emission phase has ended, but which, like the dSNRs discovered in dense molecular clouds in the Galactic disk, is expanding due to conservation of kinetic energy and momentum \citep{sof20,sof21}.
Approximating the evolution by Sedov's solution, we estimate the age to be $7\times 10^4$ y.
Therefore, although the Brick is not a star-forming region today, there has been activity in the past to form massive stars associated with a supernova explosion at its centre.

 
\section*{Acknowledgments} 
The data analysis was performed at the Astronomical Data Analysis Center of the National Astr. Obs. of Japan.
The author thanks Dr. Tomoharu Oka for the Nobeyama, Dr. H. Churchwell for Spitzer, and Drs. I. Heywood and F. Yusef-Zadeh for MeerKAT archival data. 
 
\section*{Data availability} 
The CO data were taken from
https:// www.nro.nao.ac.jp/ $\sim$nro45mrt/html/ results/data.html.
Spitzer FIR image was downloaded from:
https://irsa.ipac.caltech.edu/data/SPITZER/GLIMPSE/

\section*{Conflict of interest}
There is no conflict of interest.


\end{document}